\documentclass[twocolumn, times]{aastex631}
\usepackage[yyyymmdd,hhmmss]{datetime}
\usepackage{CJK}
\usepackage{float}
\usepackage{booktabs} 
\usepackage{xargs} 
\usepackage{chemformula}
\usepackage{subfigure}
\usepackage{xcolor, soul}
\usepackage[colorinlistoftodos,prependcaption,textsize=tiny,textwidth=.5in]{todonotes}
\newcommand{\bpic}{$\beta$ Pic}

\begin{document}

\author[0000-0001-9352-0248]{Cicero X. Lu}
\affiliation{Department of Physics and Astronomy, The Johns Hopkins University, 3400 N.~Charles Street, Baltimore, MD 21218, USA}
\email{cicerolu@jhu.edu}

\author[0000-0002-8382-0447]{Christine H. Chen}
\affiliation{Department of Physics and Astronomy, The Johns Hopkins University, 3400 N.~Charles Street, Baltimore, MD 21218, USA}
\affiliation{Space Telescope Science Institute, 3700 San Martin Dr., Baltimore, MD 21218, USA}

\author[0000-0001-9855-8261]{B. A. Sargent}
\affiliation{Department of Physics and Astronomy, The Johns Hopkins University, 3400 N.~Charles Street, Baltimore, MD 21218, USA}
\affiliation{Space Telescope Science Institute, 3700 San Martin Dr., Baltimore, MD 21218, USA}

\author[0000-0001-8302-0530]{Dan M. Watson}
\affiliation{Department of Physics and Astronomy, University of Rochester, 500 Wilson Blvd, Rochester, NY 14627, USA}

\author[0000-0002-9548-1526]{Carey M. Lisse}
\affiliation{Johns Hopkins University Applied Physics Laboratory, 11100 Johns Hopkins Rd, Laurel, MD 20723, USA}

\author[0000-0003-1665-5709]{Joel D. Green}
\affiliation{Space Telescope Science Institute, 3700 San Martin Dr., Baltimore, MD 21218, USA}

\author[0000-0003-1799-1755]{Michael L. Sitko}
\affiliation{Department of Physics, University of Cincinnati, Cincinnati, OH 45221, USA}
\affiliation{Space Science Institute, Boulder, CO 80301, USA}

\author[0000-0002-8026-0018]{Tushar Mittal}
\affiliation{MIT Department of Earth, Atmospheric, and Planetary Sciences, Green Bldg, The 77 Massachusetts Ave, Cambridge, MA 02139, USA}

\author[0000-0002-7716-6223]{V. Lebouteiller}
\affiliation{AIM, CEA, CNRS, Université Paris-Saclay, Université Paris Diderot, Sorbonne Paris Cité, F-91191 Gif-sur-Yvette, France}

\author[0000-0003-4520-1044]{G. C. Sloan}
\affiliation{Space Telescope Science Institute, 3700 San Martin
Drive, Baltimore, MD 21218, USA}
\affiliation{Department of Physics and Astronomy, University of
  North Carolina, Chapel Hill, NC 27599-3255, USA}

\author[0000-0002-4388-6417]{Isabel Rebollido}
\affiliation{Space Telescope Science Institute, 3700 San Martin Dr., Baltimore, MD 21218, USA}

\author[0000-0003-4653-6161]{Dean C. Hines}
\affiliation{Space Telescope Science Institute, 3700 San Martin Dr., Baltimore, MD 21218, USA}

\author[0000-0001-8627-0404]{Julien H. Girard}
\affiliation{Space Telescope Science Institute, 3700 San Martin Dr., Baltimore, MD 21218, USA}

\author[0000-0003-4990-189X]{Michael W. Werner}
\affiliation{Jet Propulsion Laboratory, California Institute of Technology, 4800 Oak Grove Drive, Pasadena, CA 91107, USA}

\author[0000-0002-2805-7338]{Karl R. Stapelfeldt}
\affiliation{Jet Propulsion Laboratory, California Institute of Technology, 4800 Oak Grove Drive, Pasadena, CA 91107, USA}

\author[0000-0002-5888-4836]{Winston Wu}
\affiliation{Center for Language and Speech Processing, 
Department of Computer Science, Johns Hopkins University, Baltimore, MD 21218, USA}
\affiliation{Computer Science and Engineering, University of Michigan, Ann Arbor, MI 48109, USA}

\author[0000-0002-5885-5779]{Kadin Worthen}
\affiliation{Department of Physics and Astronomy, The Johns Hopkins University, 3400 N.~Charles Street, Baltimore, MD 21218, USA}

\title{Trends in Silicates in the $\beta$ Pictoris Disk}

\shorttitle{Trends in \bpic~Silicates}
\shortauthors{Lu et al.}
\correspondingauthor{Cicero X. Lu}

\received{Nov 19, 2021}
\accepted{May 16, 2022}

\begin{abstract}
While \bpic~is known to host silicates in ring-like structures, whether the properties of these silicate dust vary with stellocentric distance remains an open question. We re-analyze the $\beta$ Pictoris debris disk spectrum from the \textit{Spitzer} Infrared Spectrograph (IRS) and a new IRTF/SpeX spectrum to investigate trends in Fe/Mg ratio, shape, and crystallinity in grains as a function of wavelength, a proxy for stellocentric distance. By analyzing a re-calibrated and re-extracted spectrum, we identify a new $18\,\mu$m forsterite emission feature and recover a $23\,\mu$m forsterite emission feature with a substantially larger line-to-continuum ratio than previously reported. We find that these prominent spectral features are primarily produced by small submicron-sized grains, which are continuously generated and replenished from planetesimal collisions in the disk and can elucidate their parent bodies' composition. We discover three trends about these small grains: as stellocentric distance increases, (1) small silicate grains become more crystalline (less amorphous), (2) they become more irregular in shape, and (3) for crystalline silicate grains, the Fe/Mg ratio decreases. Applying these trends to \bpic's planetary architecture, we find that the dust population exterior to the orbits of \bpic~b and c differs substantially in crystallinity and shape. We also find a tentative $3$--$5\,\mu$m dust excess due to spatially unresolved hot dust emission close to the star. From our findings, we infer that the surfaces of large planetesimals are more Fe-rich and collisionally-processed closer to the star but more Fe-poor and primordial farther from the star.
\end{abstract}

\section{Introduction}

Debris disks are planetary systems that contain dust, planetesimals, planets, and gas \citep{Hughes+18}, and they provide important insights into planet formation.
Theoretical models suggest that two main mechanisms efficiently remove dust grains. Stellar radiation pressure removes grains smaller than the so-called ``blowout size'' from debris disks \citep{Dent+14}, while it causes the $\mu$m-to-mm grains to drift towards their star via the Poynting-Robertson (P-R) effect \citep{Guess62}. However, debris disks are observed to be dust-rich, containing grains with a wide range of sizes from sub-$\mu$m to several millimeters in diameter. The presence of sub-blowout size grains in observations points to an active dust replenishing mechanism: collisions among parent bodies such as planetesimals, asteroids, and/or unseen planets. The \textit{Spitzer Space Telescope} has revealed signatures of such collisional activities in the mid-infrared wavelengths through spectroscopy, imaging, and time-series photometry \citep{Chen+20}.

The properties of small dust grains in debris disks, such as crystallinity and Fe-to-Mg abundance, inform us about properties of their larger parent bodies and offer a direct comparison with asteroids and Kuiper Belt Objects (KBOs) in the Solar System. 
In the Solar System, asteroidal and cometary relic planetesimals are abundant with crystalline silicates \citep{Lisse+06, Lisse+07Comp, Brownlee+08, Reach+10, Wooden+17}. 
Specifically, comets that originate from the Trans-Neptunian region contain Mg-rich silicates \citep{Wooden+17}, whereas asteroids and chondrites originating from the asteroid belt are Fe-rich \citep{LeGuillou+15}.
Similar to the Solar System, a significant fraction ($\sim 25\%$) of debris disks are also found to contain crystalline silicate grains \citep{Chen+14, Mittal+15}. However, most of these disks are too far to be spatially resolved, and thus we cannot map the crystallinity or Fe-to-Mg ratio in these disks.

As one of the few nearby ($20$\,pc) systems that can be spatially resolved by existing telescopes, \bpic~provides us with the opportunity to compare its dust distributions and properties with Solar System dust grain distributions and properties. $\beta$ Pictoris (\bpic) is a 20 Myr old A-type dwarf star and hosts dust, planetesimals, gas and at least two planets \citep{Lagrange+09, Lagrange+10, Lagrange+20, Nowak+20}. Ground-based mid-infrared spectra and images of the \bpic~disk have revealed mid-infrared (MIR) thermal emission attributed to amorphous and crystalline silicate species \citep{Weinberger+03, Telesco+05}. These silicate species display distinct spatial structures \citep{Okamoto+04}. Specifically, high angular resolution spectroscopy with Subaru COMICS has shown that the crystalline silicates are located towards the center of the disk, and sub-$\mu$m-sized amorphous silicates are distributed in three concentric rings \citep[][]{Okamoto+04,Wahhaj+03}. However, ground-based MIR observations are inherently limited by the sky thermal background and are unable to resolve regions beyond $30$\,AU from the star, missing the majority of the disk that spans out to $1400$\,AU in the \textit{Spitzer} Multiband Imaging Photometer (MIPS) $24\,\mu$m image \citep{Ballering+16} and $1800$\,AU in the scattered light image \citep{Larwood+01}.

Space-based mid and far-infrared (FIR) observations enable both the discovery and characterization of cool crystalline silicates in the far out regions of the \bpic~disk that are inaccessible to ground-based observations. 
The \textit{Spitzer} Infrared Spectrograph (IRS) has detected forsterite emission bands at $28$ and $33.5\,\mu$m, indicating a cool crystalline silicate population \citep{Chen+07}. The \textit{Herschel} Space Observatory's FIR Photodetector Array Camera and Spectrometer (PACS) has revealed a separate population of cool crystalline forsterite through the $69$\,$\mu$m forsterite band. These silicates has been found to have a Mg/Fe ratio of $99:1$ \citep{deVries+12}.  Such a Mg-rich silicate grain composition suggests that the parent bodies, planetesimals are primitive and unprocessed, similar to the comets seen in the Kuiper Belt in our Solar System. 

Since the \bpic~debris disk contains multiple populations of silicates, our goal is to investigate whether there are any trends in dust properties as a function of wavelength. In this work, we re-extract the \bpic~\textit{Spitzer} IRS spectrum using Advanced Optimal extraction (AdOpt) \citep{Lebouteiller+10} and re-analyze the silicate properties in the spectrum. In addition, we measure the $0.7$--$3\,\mu$m spectrum with the NASA Infrared Telescope facility (IRTF) SpeX to better constrain the stellar properties. In section \ref{section:observations}, we describe the IRTF observations and the re-reduction of the \textit{Spitzer} IRS observations. In section \ref{section:analysis}, we present our photosphere modeling. In section \ref{section:fitting}, we describe our modeling of the photosphere-subtracted thermal continuum and silicate emission features in detail. In Section \ref{section:abundance}, we analyze the abundance of dust grain species and the trends in grain properties. 
In Section \ref{section:discussion}, we discuss the implications of trends in grain properties. We conclude our paper in Section \ref{section:conclusion}. 

\section{Observations}\label{section:observations}

\subsection{SPEX}\label{section:SPEX}

We observe the $\beta$ Pic system using NASA's Infrared Telescope Facility (IRTF) Medium Resolution Spectrograph and Imager (SpeX) in its Short wavelength cross-dispersed (SXD) (R$\sim 2000$, $0.7$--$2.55$\,$\mu$m) and Long wavelength cross-dispersed (LXD) (R$\sim2500$,  $1.7$--$5.5$\,$\mu$m) modes \citep{Rayner+2003} on 2021 February 02 at \formattime{05}{47}{18} UT and \formattime{07}{12}{12} UT. We use nearby (within $0.5$~deg and $0.1$ in airmass) HD 37781 (A0V, $K~=\,6.516$) as a calibration standard to measure the atmospheric effects on our observations of \bpic. We observe $\beta$ Pic and HD 37781 in SXD mode with a total on-target integration time of $92$s and $180$~s, and in LXD mode for a total on-target integration time of $66$~s and $637$~s respectively, at an airmass $\sim$3. Both stars are observed in ABBA nod mode for telescope and sky background removal.

We reduce our data using Spextool v4.1 \citep{Cushing+04, Vacca+03}. The calibrated \bpic~SXD spectrum has a signal-to-noise ratio (SNR) of $100$--$300$, similar to that of the calibrator HD 37781. However, although the \bpic~LXD has SNR of $100$--$300$, most regions in the calibrator's LXD has SNR of $<1$, due to insufficient on-target integration time (calibrator is $\sim4$ mags fainter than \bpic). As a result, the calibrated \bpic~spectrum has SNR$\ge~10$ only at wavelengths $1.7$--$2.55$ and $3.0$--$4.0$\,$\mu$m. The region in between these windows is heavily impacted by the atmospheric transmission window.

We perform absolute flux calibration of the IRTF SXD spectrum using ESO VLT/NACO JHK photometry \citep{Bonnefoy+13}. First, we calculate the synthetic photometry $F^{\text{SXD}}_J$, $F^{\text{SXD}}_H$, and $F^{\text{SXD}}_K$ by convolving the SXD spectrum with NACO JHK's filter transmission functions. Next, we calculate a scaling factor 
\begin{equation}
    C_{\text{SXD}} = \frac{ F^{\text{SXD}}_J + F^{\text{SXD}}_H + F^{\text{SXD}}_K   }{   F^{\text{NACO}}_J + F^{\text{NACO}}_H + F^{\text{NACO}}_K   }.
\end{equation}
We calculate a scaling factor $C_{\text{SXD}}$ of $1.04$, indicating that the $F^{\text{SXD}}$ for JHK bands are on average $4$\% dimmer than $F^{\text{NACO}}$ for JHK bands. This $4$\% difference is within the IRTF SpeX instrumental accuracy \citep[$5\%$, ][]{Rayner+2003}. Therefore, we multiply the IRTF SpeX spectrum by $1.04$ to be consistent with NACO photometry data. The SNR of the LXD spectrum is limited by the SNR of the standard star and thus suffers from a $\sim${}30\% loss in brightness compared to the L' band photometry. In addition, the LXD spectrum is heavily impacted by the atmospheric transmission. Therefore, we do not perform any further analysis on the LXD spectrum.

\subsection{\textit{Spitzer} IRS}

We re-extract and re-calibrate the \emph{Spitzer} IRS \citep{Houck+04} low-resolution  \bpic spectrum \citep[originally published in ][]{Chen+07} with the most up-to-date IRS extraction and calibration tools \citep{Lebouteiller+10}. In the original \bpic~spectrum, \citet{Chen+07} discovered the $23$ and $35\,\mu m$ crystalline silicate emission bands. However, their spectrum displays ``sawtooth" fringing patterns as a result of detector artifacts that were not able to be corrected for at the time.  With advancements in both \textit{Spitzer} science center pipelines and knowledge of empirical Point Spread Functions (PSFs), we now attempt to minimize those detector artifacts that could mask astrophysical signals. In the following subsections, we describe in detail our procedures for re-extracting and re-calibrating the \bpic~observations.

IRS is a MIR spectrograph that covers $5.2-38\,\mu m$ with low ($R\sim60-130$) and moderate ($R\sim600$) resolution spectroscopic capabilities, and has two operating modes: a mapping mode and a staring mode. For our work, we focus on low-resolution observations of \bpic. The low resolution mode consists of two modules, Short-Low (SL) and Long-Low (LL). Both have three grating orders with two main orders SL1, SL2, LL1, LL2 and a short ``bonus'' order (SL3, LL3). Since SL3 (LL3) is observed simultaneously with the SL1 (LL1) and shares the same observation setups, we omit a separate description of SL3 (LL3). \bpic~is observed with a combination of mapping mode and staring mode. See Table \ref{tbl:obs_setup} for details. All low resolution observations are made with the spectrograph long slit aligned along the position angle of \bpic~disk to within $15^{\circ}$. We describe details in the following subsections.

\begin{deluxetable*}{lcccccccc}
    \tablecaption{Observation setup}\label{tbl:obs_setup}
    \tablehead{
        \colhead{Order} &
        \colhead{Wavelength} &
        \colhead{Mode} &
        \colhead{Date} & 
        \colhead{AOR Key} & 
        \colhead{\# Pointings} &
        \colhead{Pointing Extracted} & 
        \colhead{Slit Size} & 
        \colhead{Plate scale} \\ 
        \colhead{} & \colhead{($\mu$m)} & \colhead{}  & \colhead{} & \colhead{} & \colhead{} &\colhead{}&\colhead{} &\colhead{(''/pix)}
    }
    \startdata
     SL2 & $5.2$--$7.7$   & Mapping & 2004 Nov 16 & 9872288 & 7  & Exp $4$ & $57\arcsec \times 3.7\arcsec$ & 1.8\\
     SL3 & $7.3$--$8.7$   & Mapping & 2004 Nov 16 & 8972544 & 11 & Exp $6$ & $57\arcsec \times 3.6\arcsec$ & 1.8\\
     SL1 & $7.4$--$14.5$  & Mapping & 2004 Nov 16 & 9872288 & 11 & Exp $6$ & $57\arcsec \times 3.6\arcsec$ & 1.8\\
     LL2 & $13.9$--$21.3$ & Mapping & 2005 Feb 9  & 9016832 & 5  & Exp $3$ & $168\arcsec \times 10.5\arcsec$ & 5.1\\
     LL3 & $19.4$--$21.7$ & Staring & 2005 Feb 9  & 9016832 & 2  & Exp $1$, $2$ & $168\arcsec \times 10.7\arcsec$  & 5.1\\
     LL1 & $19.9$--$38.0$ & Staring & 2004 Feb 4  & 9016064 & 2  & Exp $1$, $2$ & $168\arcsec \times 10.7\arcsec$  & 5.1
    \enddata
    \tablecomments{``SL'' stands for Short-Low module and ``LL'' stands for Long-Low module. See Figure \ref{fig:IRSmappingPattern} for a visualization of mapping mode dither patterns. }
\end{deluxetable*}

\subsubsection{\bpic~Observations using the IRS Mapping Mode}

\begin{figure}[h!]
    \epsscale{1}
    \plotone{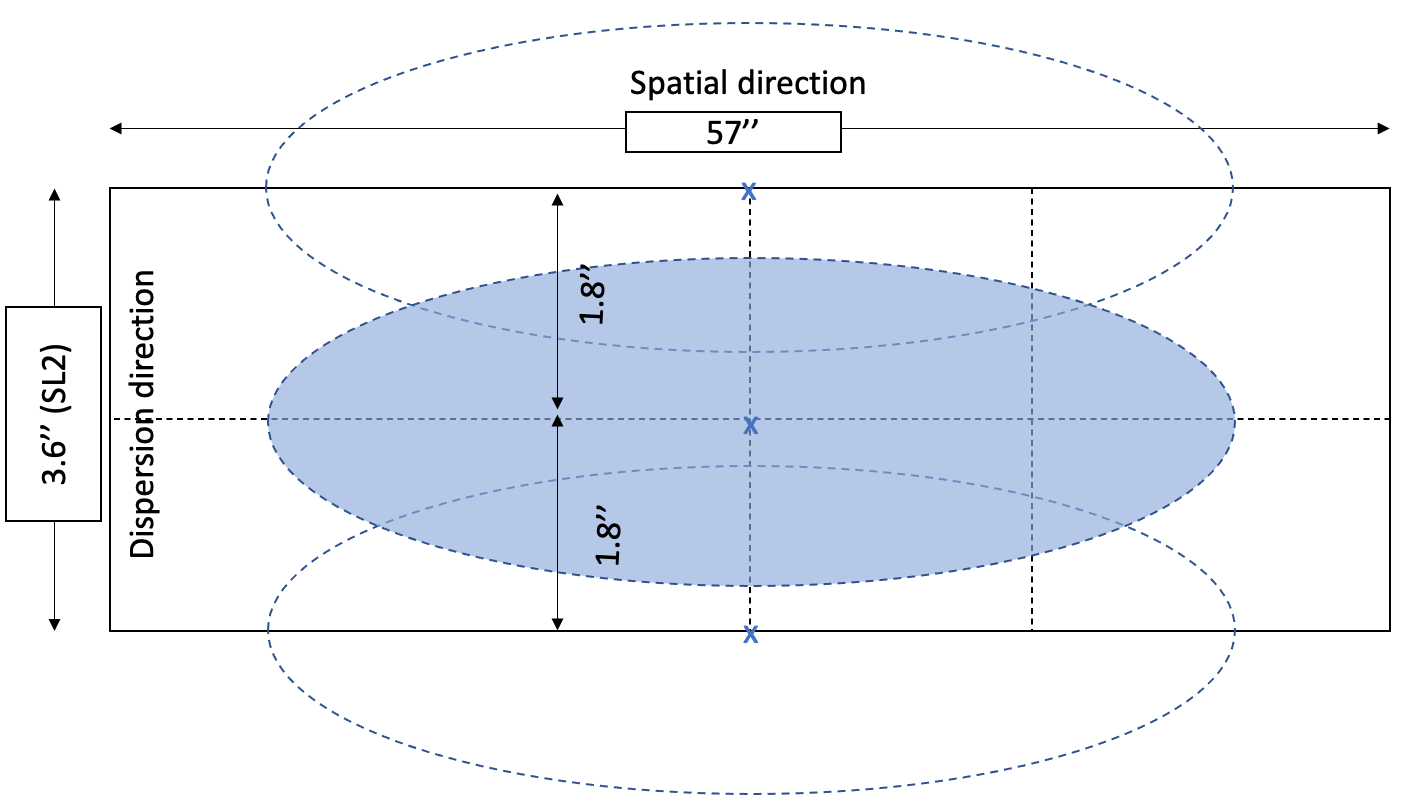}
    \caption{A cartoon illustration of the position of the \bpic~disk with respect to the SL slit ($57\arcsec \times 3.6\arcsec$ slit) for the IRS SL1 and SL2 mapping mode observations. The telescope was stepped across the \bpic~disk in the dispersion direction in increments of 1.8$\arcsec$ with the disk well-centered in the central pointing as illustrated by the shaded ellipse. The pointings represented by the dotted empty ellipses are not centered in the dispersion direction. The LL2 observations follow similar mapping patterns, where only the central pointing is well-centered in the dispersion direction. Drawing not to scale. 
    }
    \label{fig:IRSmappingPattern}
\end{figure}

During SL2, SL3, SL1, and LL2 observations, \textit{Spitzer} mapped the vertical extent of the disk by stepping the telescope across the disk at a specific number of positions (\# Pointings in Table \ref{tbl:obs_setup}), each separated by $1.8\arcsec$ except for LL2 ($2.1\arcsec$ for LL2). Because the slits are long and narrow (Slit Size), the disk is only fully captured in the slit in the central map position (Pointing Extracted), where the star is well centered in the slit in the dispersion direction. Figure \ref{fig:IRSmappingPattern} shows a cartoon of three SL2 pointings including the central pointing (exposure 6). In the rest of the pointings, the midplane of the disk is outside of the slit and results in low SNR. Thus, we only perform analysis on the central pointing.

For well-centered exposures, we compare slit sizes with the size of \bpic~disk and determine the effect of telescope PSF on SL and LL observations to ensure that the disk is completely captured in the slit. Since \textit{Spitzer} is diffraction-limited, the PSF scales as $1.22\,\lambda/D$, where D is the diameter of Spitzer's mirror and $\lambda$ is the wavelength.  For SL observations, the PSF is  $~1.5\arcsec$ ($1$ pixel) at $5\,\mu m$ and  $~4.4\arcsec$ ($2.5$ pixel) at $15\,\mu m$. For LL observations, the PSF is $~4.4\arcsec$ ($0.9$ pixel) at $15\,\mu m$ and $10.4\arcsec$ ($2$ pixel)  at $35\,\mu m$. We find that the \bpic~disk roughly spans $3$ pixels (approximately $15\arcsec$) $15\,\mu m$ in the radial direction and therefore the slits are long enough to capture the entire disk. We also find that disk is spatially resolved in both SL and LL and plan to publish the spatially-resolved IRS spectra of the \bpic~disk in a subsequent paper.  We measure the misalignment angle between the LL slit position angle ($\text{PA}_{\text{slit}}$ = $44.97 \pm 0.02^\circ$) and disk midplane PA ($\text{PA}_{\text{disk}}= 29.51^\circ$) to be  $15^\circ$. We overlay the IRS LL2 slit on top of the data for a visualization (see Appendix Figure \ref{fig:diskPA}). We conclude that the minor misalignment does not affect the observation such that the entire disk is captured in the well-centered exposures. 

We re-extract SL1,2,3 and LL2 spectra with Advanced Optimal extraction \citep[AdOpt,][]{Lebouteiller+10}.
AdOpt first uses empirical super-sampled PSFs to simultaneously fit the PSF to all of the pixels in the spatial direction of the slit and determine weights for individual pixels. In doing so, AdOpt weights the pixels in the extraction window by their SNR and position on the detector to measure the flux at every wavelength. We do not observe any obvious effects of fringing in the extracted SL1,2,3, and LL2 spectra and therefore do not apply fringing correction. 

\subsubsection{\bpic~Observations using the IRS Staring Mode}

The \bpic~LL1 and LL3 spectra were observed with IRS staring mode, which stepped the disk at two nod positions (equivalent to pointings) at one-third (Nod 1) and two-thirds (Nod 2) along the slit. First, we extract the spectrum from each nod position with AdOpt. \bpic~is well-centered in the dispersion direction of the slit in Nod 1 but not well-centered in Nod 2. The flux in Nod 2 is $20$\% lower across all wavelengths than in Nod 1. For the IRS staring mode, the AdOpt supports user-input extraction positions. Therefore, we use the manual optimal extraction option in SMART \citep{Lebouteiller+10} to adjust the extraction position on the detector plane. We test offset position values between $-1$ and $+1$ in $0.1$ increments. We find that an extraction offset (offset = $-0.39$ pixels) best matches the radial profile for the detector image by minimizing the residual, (the difference between between the data and extraction profile) across all LL1 wavelengths. After manually extracting the Nod 2 spectrum, we find that the flux between Nod 1 and Nod 2 spectra are consistent with one another to within $\leq 1$\%. 

We find artifacts from fringing in LL1 spectra and corrected for them by using an empirical Relative Spectral Response Function (RSRF) to correct for out-of-slit light losses. The calibrator star's RSRF spectrum for is defined as the model photosphere divided by the empirical spectrum and therefore characterizes the detector artifacts across pixels. We construct a one-dimensional RSRF for the LL1 wavelength range using a standard K giant star, $\xi$ Dra. $\xi$ Dra was observed on 2005 February 12 (AOR key 13195008) as a part of the IRS calibration program \citep{Sloan+15}. Since $\xi$ Dra is very bright in the mid IR ($7$ Jy at $15\,\mu m$) and has no observed infrared excess, its spectrum is approximately a bare stellar photosphere. We divide out a normalized $\xi$ Dra IRS spectrum from the \bpic~spectrum for the RSRF correction. The fringe-correction is effective in removing the fringing effects (e.g., bumps and wiggles) from the spectrum. 

\subsection{Absolute Flux Calibration and Order Stitching}
We perform absolute flux calibration by pining the LL1 spectrum to the MIPS $24\mu m$ flux using LL1 as an anchoring order to calibrate the rest of LL and SL spectra. 
First, we perform our own MIPS $24\mu m$ aperture photometry extraction by measuring the flux of the unresolved point source in the MIPS $24\,\mu m$ image. We cannot use existing MIPS $24\mu m$ photometry reported in \citet{Ballering+16} as AdOpt's extraction window differs from that used by \citet{Ballering+16}. AdOpt weights the extraction for the pixels by their SNR and their relative position on the detector. In doing so, AdOpt emphasizes the contribution from the high SNR unresolved point source. Specifically, we calculate the MIPS $24\,\mu m$ flux, $F_{24\,\mu m, \text{MIPS}}$ that is consistent with an unresolved point source, using aperture photometry with a radius of $3.5$\arcsec. We use the IDL-based tool, Image Display Paradigm \#3 (IDP3) \citep{Lytle+99} with background subtraction and aperture correction. We choose an annulus with an inner radius of $30$\arcsec and an outer radius of $30.5$\arcsec from disk center as the background, because the disk attenuates out to roughly $30$\arcsec \citep{Ballering+16}. Next, we multiply our extracted flux by $2.57$, the aperture correction given in the MIPS Instrument Handbook for a $3.5$\arcsec aperture. We estimate $F_{24\,\mu m, \text{IRS}}$, the synthetic photometry from the IRS spectrum. We convolve the MIPS $24\,\mu m$ filter response function with the IRS spectrum. For the unresolved central point source, we estimate the $F_{24\,\mu m, \text{MIPS}}$ = $6.66$~Jy and the $F_{24\,\mu m, \text{IRS}} = 7.01$~Jy. Therefore, we apply a scaling factor  $C_{\text{IRS}} = F_{24\,\mu m, \text{IRS}}/F_{24\,\mu m, \text{MIPS}} = 0.95$ to the IRS observations to make them consistent with the MIPS $24\,\mu m$ observations. We report this scaling factor and any subsequent ones in the ``Scaling Factor'' column in Table \ref{tbl:absFlux} and photometry data in Table \ref{tbl-5}.

We use flux-calibrated LL1 spectrum  as an anchoring spectrum to scale the rest of LL and SL spectra in descending wavelength order. There is an overlapping wavelength range (column ``Overlapping Wavelength'' in Table \ref{tbl:absFlux}) between every two adjacent orders (column ``Ref. Order or Photometry'' in Table \ref{tbl:absFlux}).  To calibrate the flux in an order, we take the data points in its overlapping wavelength range with its reference order and calculate an average flux, $f_{\text{order}}$. We repeat this procedure for its reference order and obtain $f_{\text{ref. order}}$. Then, we take the ratio of the two to be the scaling factor $C_{order} = f_{\text{ref. order}}/f_{\text{order}} $. Specifically, take LL3 for an example, LL3 and LL1 overlaps between $19.9$--$21.7\,\mu m$. The scaling factor is $C_{LL3} = \sum_{19.9 \mu m}^{21.7 \mu m} f_{LL1}(\lambda)/\sum_{19.9 \mu m}^{21.7 \mu m} f_{LL3}(\lambda) = 1.12 $. We report the rest of scaling factors in Table \ref{tbl:absFlux}. Finally, we check our absolute flux calibration for the entire IRS spectrum with WISE photometry and we show that our absolute flux calibration is consistent with WISE in Fig. \ref{fig:Photosphere}.

\begin{deluxetable*}{lccccc}
    \tablecaption{Absolute Flux Calibration Parameters}\label{tbl:absFlux}
    \tablehead{
        \colhead{Order} &
        \colhead{Wavelength} &
        \colhead{Ref. Order or Photometry} &
        \colhead{Overlapping Wavelength} &
        \colhead{Scaling Factor} \\ 
        \colhead{} & \colhead{($\mu$m)} & \colhead{} & \colhead{} & \colhead{($C_{order}$)}
    }
    \startdata
     LL1 & $19.9$--$38.0$ & MIPS24 & \nodata & $0.95$\\
    LL3 & $19.4$--$21.7$ & LL1  & $19.9$--$21.7$ & $1.12$\\
     LL2 & $13.9$--$21.3$ & LL3  & $19.4$--$21.3$   & $1.18$\\
     SL1 & $7.4$--$14.5$  & LL2  & $13.9$--$14.5$ & $1.35$\\
     SL3 & $7.3$--$8.7$   & SL1  & $7.4$--$8.7$ & $1.0$\\
     SL2 & $5.2$--$7.7$   & SL3  & $7.3$--$7.7$  & $1.07$
    \enddata
\end{deluxetable*}

\subsubsection{\textit{IRS} spectrum Uncertainties}

\begin{figure*}[]
\epsscale{0.9}
\plotone{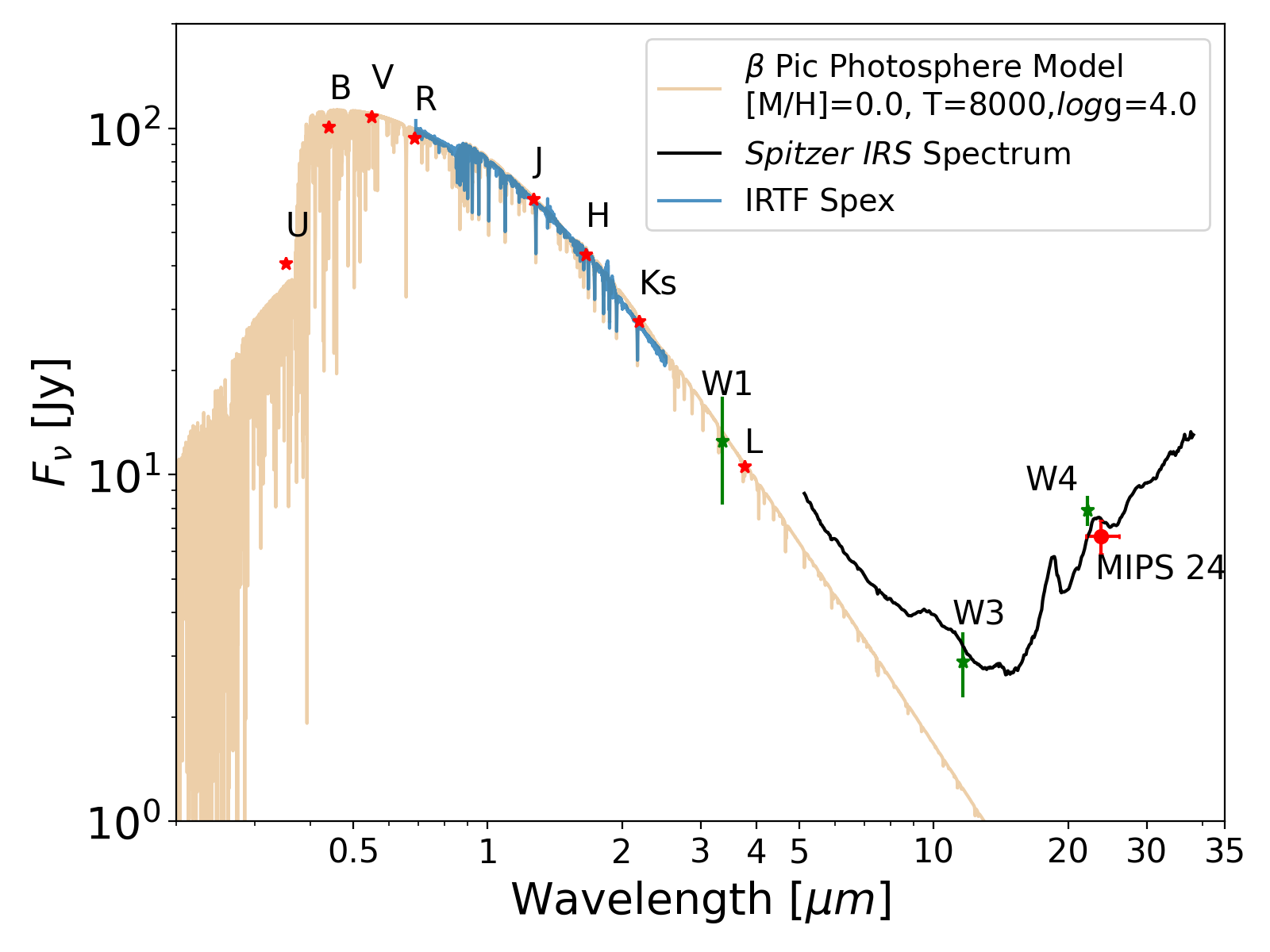}
\caption{\bpic~SED showing our best-fit model for the stellar photosphere overlaid on UBVRJHK$_s$ photometry (see Table \ref{tbl-photometry}) and our $0.7$--$2.5$\,$\mu$m IRTF spectrum. Our best-fit model has [M/H] = $0.0$, T${_\text{eff}} = 8000 K$, $\log g=4$, $A_v = 0.078$ and $R_V$ = 3.1. The red stars indicate photometry used in the stellar photosphere fitting, and the red circle is the MIPS $24\,\mu m$ flux derived here for the unresolved central point source. The horizontal error bars indicate the FWHM of the band width for the MIPS $24 \mu m$ filter. The L band photometry indicates that the disk has no infrared excess at $3.77\,\mu$m.  We omit W2 photometry because it is unreliable for sources brighter than magnitude of 5. \label{fig:Photosphere}}
\end{figure*}

We take separate approaches to determine uncertainties for staring mode and mapping mode observations of the $\beta$ Pic. For orders observed in IRS Staring mode (LL1 and LL3), we take the absolute value of the difference in flux between the two nod positions as the uncertainty of the spectra. For orders observed in IRS Mapping mode (SL1, SL2 and LL2 orders), we fit polynomials to part of the spectrum that are not affected by solid state features. We select regions at $5.6$--$7.9$ and  $14.32$--$14.83$\,$\mu$m  of the spectrum and measure the root mean square (rms) of the spectrum from the polynomial fit. We assign the rms as the uncertainty for the spectrum if the rms is bigger than $1$\% and adopted a 1\% error floor according to \citet{Higdon+04}. The resulting spectrum is shown in Fig. \ref{fig:Photosphere}.

\section{Analysis}\label{section:analysis}

In this section, we first describe fitting our new IRTF spectrum and existing photometry with stellar photosphere models to better predict the stellar photospheric emission at mid-infrared wavelengths. We then report our discovery of new silicate emission features from AdOpt extraction. Lastly, we describe our discovery of an infrared excess at $3$--$5$\,$\mu$m, consistent with the presence of hot dust in the system. 

\subsection{Stellar Photosphere}

We model the \bpic~stellar photosphere to understand (1) the relative flux contribution of disk emission to the overall brightness at the shortest wavelength in IRS \bpic~spectrum and (2) to rigorously determine the overall shape of the disk emission spectrum. At H-band, Very Large Telescope Interferometer (VLTI) measurements suggest that the \bpic~disk emission only constitutes $0.88\%$ of the total emission \citep{Ertel+14}. Similarly, at the shortest wavelengths of the IRS spectrum ($5.3\,\mu$m), the spectrum is expected to be dominated by the stellar photosphere. We use the VLTI measurement to accurately estimate the brightness of the stellar photosphere at the shortest IRS wavelengths.

To predict the stellar photosphere at 5.5 $\mu$m, we fit $\beta$ Pic's UBVRJHK$_s$ photometry (Table \ref{tbl-photometry}) and IRTF spectrum from $0.7$--$2.5$\,$\mu$m with BT-NextGen models \citep{Allard+12,Hauschildt+99}. The addition of an IRTF $\beta$ Pic spectrum better constrains the model's spectral slope in the infrared wavelength range. \bpic~has an edge-on disk, and therefore the disk might provide a small amount of extinction along line-of-sight. Therefore, we assume that extinction, E(B-V), is a free parameter and redden photosphere models using the general extinction law with $R_v = 3.1$ with ``dust Extinction'' software. We also include the effect of stellar rotation and limb darkening by convolving photosphere models with a line spread function consistent with $v\sin i = 130~ \textrm{km}\cdot s^{-1}$ \citep{Claret2000} and applying a limb-darkening coefficient of 0.24 consistent with the H band measurements of \bpic~ \citep{Claret+95}. Our best-fit model has $T_{\text{eff}} = 8000 K$, $R_v$ = $3.1$, $A_v$ = $0.078$, $\log g=4.0$ and $\text{[M/H]} = 0.0$ with a reduced $\chi^2=0.012$. Our best fit values for $\log g$, metallicity and the $T_{\text{eff}}$ are consistent with those reported in the literature \citep[e.g.,][]{Pecaut+Mamajek13}. We note that this fit incorporates extinction as a free parameter for the first time. In figure \ref{fig:Photosphere},  we show the $\beta$ Pic stellar photosphere model, the IRTF spectrum and the \textit{Spitzer} IRS spectrum together. 

\begin{figure}
\epsscale{1.2}
\plotone{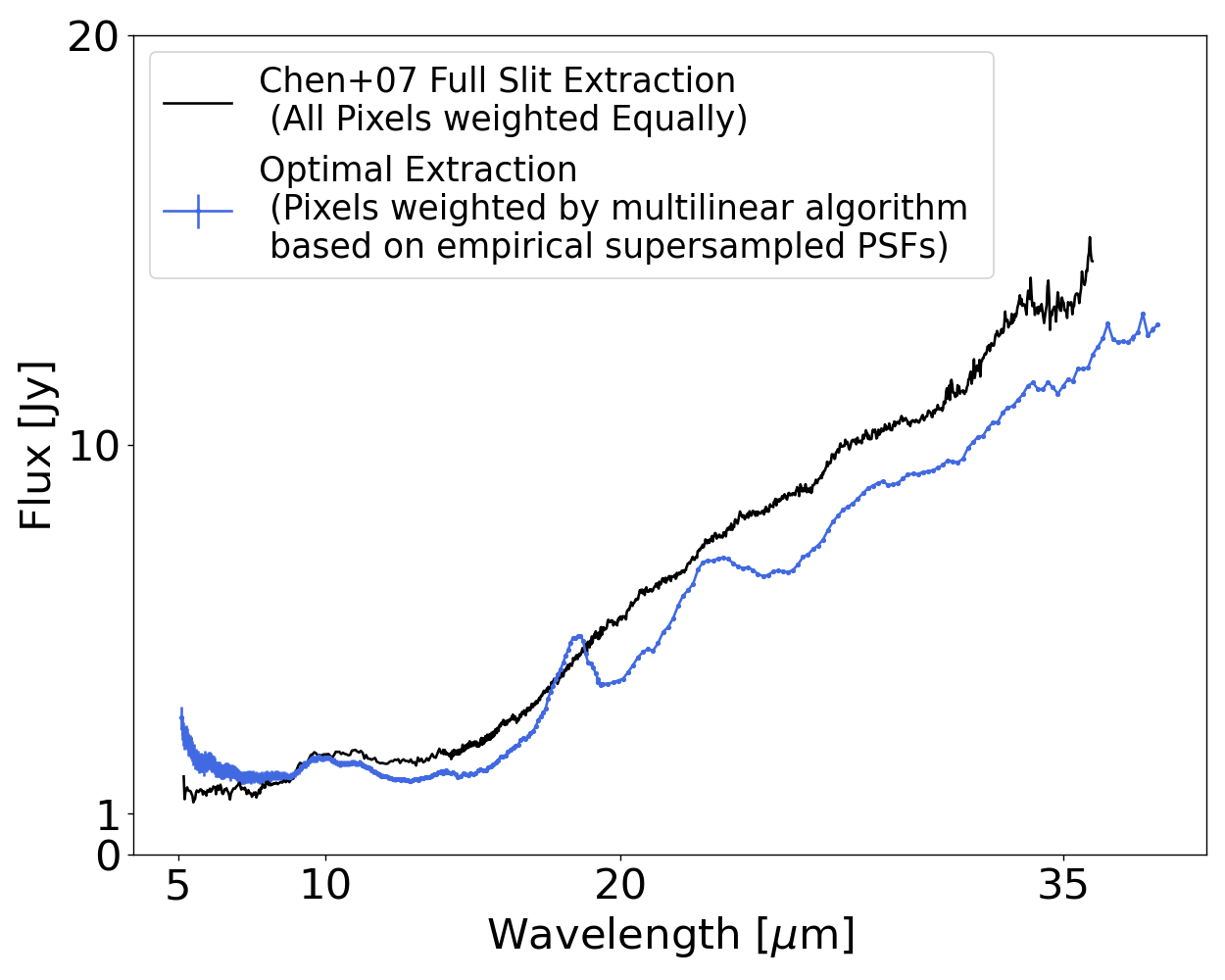}
\caption{The $\beta$ Pic IRS spectrum showcasing the newly discovered $18\,\mu$m feature and a $23\,\mu$m feature with larger line-to-continuum ratio that previously reported by \citet{Chen+07}. The optimal extraction (this work) is in blue, while the full slit extraction published in \citet{Chen+07} is in black. The uncertainties for the spectrum extracted with AdOpt are on average at $1\%$ level. The Chen+07 spectrum reduced using a full slit extraction has fringes. \label{fig:compare}}
\end{figure}

Next, we subtract off our best-fit stellar photosphere model from the IRS AdOpt spectrum. In figure \ref{fig:compare},  we plot the photosphere subtracted AdOpt spectrum of the unresolved point source in blue and the \citet{Chen+07} full-slit extraction spectrum in black for comparison. We can see the change in extraction window sizes brings out a new spectral feature at $18.5\,\mu$m and recovers the $23.7\,\mu$m crystalline forsterite feature previous reported in \citet{Chen+07} with a higher line-to-continuum ratio.

\begin{deluxetable*}{lcccc}
\tablecaption{Photometric Measurements of \bpic}\label{tbl-5}
\tablehead{
\colhead{Filter} & \colhead{Effective Wavelength Midpoint
} & \colhead{Magnitude} & \colhead{Flux} & 
\colhead{Reference}\\
\colhead{} & \colhead{$\lambda_{\text{eff}}$ for standard filters} &\colhead{} &\colhead{} &
\colhead{}\\
\colhead{} & \colhead{($\mu m$)} &\colhead{(mag)} &\colhead{(Jy)} &
\colhead{}}
\startdata
U & $0.3518$ & $4.13\pm 0.01$  & $ 40.6\pm 0.4$ & (a)\\
B & $0.4407$ & $4.03 \pm 0.01$ & $ 100.9\pm 1.0$ & (a)\\
V & $0.5479$ & $3.86 \pm 0.01$ & $ 108.0\pm 1.1$ & (a)\\
R & $0.6864$ & $3.74\pm 0.03$  & $ 93.4\pm 2.6$ & (b)\\
J & $1.265$ & $3.524 \pm 0.009$ & $ 62.4\pm 0.75$ & (c)\\
H & $1.66$ & $3.491 \pm 0.009$ & $ 43.2\pm 0.4$ & (c)\\
$K_s$ & $2.18$ & $3.451 \pm 0.009$ & $27.8 \pm 0.3$ & (c)\\
L & $3.77$ & $ 3.454 \pm 0.03$ & $ 10.6 \pm 0.03$ & (d)\\
W1 & $3.35$ & $ 3.484 \pm 0.4$ & $ 12.5 \pm 4.3$ & (e)\\
W3 & $11.6$ & $ 2.597 \pm 0.3$ & $ 2.9 \pm 0.6$ & (e)\\
W4 & $22.1$ & $ 0.014 \pm 0.016$ & $ 7.9 \pm 0.8$ & (f)\\
MIPS24 (Full Disk) & $23.7$ & \nodata & $ 7.28 \pm 0.73$ & (g)\\
IRS 24 (Unresolved Point Source ) & $23.7$ & \nodata & $ 7.01 \pm 0.7$ & This Work\\
MIPS24 (Unresolved Point Source) & $23.7$ & \nodata & $ 6.66 \pm 0.67$ & This Work
\enddata
\tablecomments{(a). The General Catalogue of Photometric Data (GCPD) \citet{Mermilliod+97} (b). \citet{Ducati02} (c). \citet{Bonnefoy+13} (d). \citet{Bouchet+91} (e). AllWISE Source Catalog \citet{Wright+10,Mainzer+11, Cutri+12}. Note: W1 and W2 have $24$\% pixels saturated and W3 has $8$\% of pixels saturated. The saturation affect measurements precision as reflected in the inflated errorbars. (f). \citet{Morales+12} (g). \citet{Su+06, Ballering+16} \label{tbl-photometry}. 
We exclude W2 photometry because for source brighter than magnitude of 5, W2 photometry becomes unreliable.}
\end{deluxetable*}

\subsection{Discovery of the new spectral features}
\label{subsection: T-ReCS}
We discover an $18.5\,\mu$m  spectral feature. We attribute the discovery of a new  $18.5\,\mu$m feature to advancements in the knowledge of empirical \textit{Spitzer} PSFs \citep{Sloan+15, Lebouteiller+10}. These improvements enable us to (1) extract a spectrum with high SNR in the slit and (2) correct for fringing in the spectrum to further validate the fidelity of the new spectral feature using all of the calibration data obtained during the cryogenic mission. Specifically, compared to the \citet{Chen+07}'s full slit extraction, which weights every pixel in the entire slit equally, our AdOpt spectrum emphasizes the region close to the star. As the SNR increases towards the central star, most of the data points in our spectrum has less than 1\% uncertainties.

We conclude that $18 \mu m$ feature must be astrophysical and is emitted by dust grains in the disk. For the newly discovered $18 \mu m$ feature, we verify that it is not a detector artifact by examining the entire IRS calibrator star library for LL2 observations to understand detector characteristics. Most fringing patterns only spans only $2$ to a few (usually $5$) data points but our $18 \mu m$ feature has a FWHM of $\sim 2 \mu m$ that spans more than $30$ data points. More importantly, we do not see any $18 \mu m$ artifacts in the calibrator star spectra that resemble our new $18 \mu m$ feature. Therefore, we rule out the possibility that the $18 \mu m$ arise from detector artifacts.

\subsection{Constraints on the Spatial Distribution of the New $18 \mu m$ Feature}
Next, to constrain the spatial distribution of dust grains that are responsible the $18 \mu$m and $23 \mu$m spectral features, we analyze the Gemini Thermal-Region Camera Spectrograph (T-ReCS) spatially-resolved broadband MIR images of \bpic~disk \citep{Telesco+05}. \citet{Telesco+05} took image of the \bpic~disk in Qa (central wavelength at $18.3\mu$m) and Qb (central wavelength at $24.6 \mu$m) bands. In these broadband MIR images, both the disk continuum emission and the characteristic solid-state emission from dust contribute to the flux. As Qa and Qb bandpass's wavelength range ($17.57$--$19.08$ and $23.62$--$25.54\,\mu m$) overlaps with our IRS solid-state emission features ($18.5\,\mu$m and $23.8\,\mu$m), these MIR images can constrain the spatial distributions of the dust grains responsible for the solid state features. The T-ReCS images are diffraction-limited with beam sizes of $0.356\arcsec$ ($7$ AU) at $18.3 \mu$m and $0.445\arcsec$ ($9$ AU) at $24.6 \mu$m, respectively, much finer than that of the Spitzer IRS. Therefore to include the effect of the changing PSF size with wavelength, we convolve the $18.3 \mu$m image with the a PSF profile at $24.6 \mu$m, such that when we take the ratio of two images in subsequent analysis, the PSF will not bias the results. 

\begin{figure}
\epsscale{1.2}
\plotone{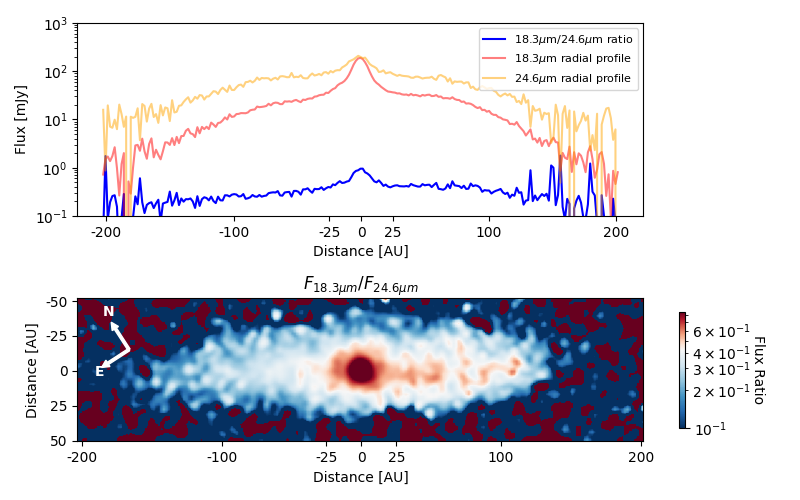}
 \caption{Top: Radial surface brightness profile of the $18.3\,\mu m$ band and $24.6\,\mu m$ \bpic~images from T-ReCS \citet{Telesco+05} imaging and ratio of the surface brightness profile. Bottom: Flux ratio between the $18.3\,\mu m$ band and $24.6\,\mu m$ from T-ReCS imaging \citep{Telesco+05}. The red color indicates regions in which the $18.3\,\mu m$ emission is relatively bright compared with the $24.6\,\mu m$ emission. The bottom panel shows that the 18 micron flux originates from regions (red colors) close to central star (within $10$'')  while the 24 $\mu$m flux originates from more distant regions in the disk (blue).  \label{fig:TRECS}}
\end{figure}

We find that majority of the emission at $18.3\,\mu m$ comes from region within $2\arcsec$ ($\sim50$ AU) and that $18\,\mu m$ emission arises from a spatial extent closer to the star than that of the $24\,\mu m$ emission. As shown in Figure \ref{fig:TRECS}, we construct a 1-D surface brightness profile of the disk at $18.3$ and $24.6\,\mu m$.  To do so, we exclude the top and bottom 10 rows of pixels to eliminate the background and then sum the flux along the y-axis direction. We find that the $18.3\,\mu m$ brightness profiles drops sharply at $\sim 2\arcsec$, indicating that the majority of the flux at $18.3\,\mu m$ comes from regions within $\sim50$ AU. We compare the spatial distribution of $18.3\,\mu m$ emission with that of $24.8\,\mu m$ by taking the ratio of the two profiles. We find that the F$_{18\,\mu m}$/ F$_{24\,\mu m}$ flux ratio also drops sharply at $\sim 2\arcsec$, indicating that the $18.3\,\mu m$ emission is mostly concentrated in the inner $\sim50$ AU, while the $24.6\,\mu m$ emission is more spread out throughout the disk. The brightness profiles shown in Figure $16$ of \citet{Ballering+16} display a similar conclusion.

We estimate the physical location of the grains responsible for the new $\sim 18\,\mu$m emission feature and compare this distance with the semi-major axis ($\sim10\,$AU) of \bpic~b. We assume that the dust is optically thin and is in radiative equilibrium. If the dust grains are large, then they will absorb and emit radiation like black bodies and be located at the black body distance. For example, large dust grains at a distance of 9 AU are expected to have a temperature of $160\,$K and to emit black body radiation whose emission peaks at $\sim 18\,\mu$m. However, scattered-light images of debris disks indicate that the black body distance tends to underestimate the actual distance of dust by a factor of $\sim 2$ \citep{Schneider+18}. Therefore, the grains that are responsible for the $\sim 18\,\mu$m feature are expected to be located at $\sim 20$AU. For comparison, the \bpic~b planet is located at $10\,$AU from the star, interior to the dust that produces the feature $18\,\mu$ m. Given the loose constraints set by the black body distance and the Gemini T-ReCS observations, there is still some uncertainty in the relation of the $18\,\mu$m dust distance to the \bpic~b planet.

We exclude the possibility that the new 18 $\mu$m feature arises from a halo component. Recent modeling of multi-wavelength \bpic~debris disk images \citep[e.g.,][]{Ballering+16} requires the presence of a spatially extended halo ($45\,$AU--$1800\,$AU or $\sim\,2.25\arcsec$ to $90\arcsec$) of fine dust grains to reproduce the thermal emission SED. If the 18 $\mu$m feature were generated in the halo, then the spectral feature would have been present in the full slit extraction of the beta Pic spectrum \citep{Chen+07}. However, the full slit extraction does not show an 18 $\mu$m  micron feature. Our optimal extraction, on the other hand, weights pixel fluxes by their SNRs in the extraction. In doing so, the extraction heavily weights the emission from the unresolved point source. Specifically,  an unresolved point source is expected to have an FWHM of $\sim\,1.5\arcsec$ at $5\,\mu m$ micron and $\sim\,7\arcsec$ at $24\,\mu m$ ($\sim\,5\arcsec$ at $18\,\mu m$). Indeed, at $5$--$7.6\,\mu m$, the AdOpt extraction window is too small to include the halo; therefore, we rule out the halo contribution to the spectrum at these wavelengths. At longer wavelengths, the extraction window includes increasingly more of the halo until it reaches a maximum FWHM $\sim\,12\arcsec$ at $40\,\mu m$ . Even at this longest wavelength, the extraction aperture is not large enough to capture all of the halo flux given the halo geometry from \citet{Ballering+16}. Since the full slit extraction did not reveal the new 18 $\mu$m feature , we conclude that the new $18\,\mu$m feature in our optimal extraction spectrum is not due to the halo component in the \bpic~disk.

\subsection{Tentative Evidence of Weak Infrared Excess around $3$--$5\,\mu$m}\label{sub:hotdust}
In the $5$--$7\,\mu$m region, the \textit{IRS} spectrum is above the stellar photosphere model, indicating a possible excess at $5\,\mu$m. The elevated flux in the spectrum is unlikely to be an artifact because the uncertainty in the point-to-point calibration of \textit{IRS} spectra is less than $1$\%. This $5 \mu$m excess is tentative evidence for the existence of a hot dust population at $\sim600\,K$ which must be physically located within $0.7$\,AU to the star.  A 5 $\mu$m excess has not been previously reported; however, near infrared interferometric observations have discovered a $0.88$\% excess at H band \citep{Defrere+12,Ertel+14}. We used all of the available photometric measurements including WISE W1, W3, W4, and an L band measurement from \citet{Bouchet+91} along with our new IRTF SPEX spectra to determine the onset of the near-infrared excess. As shown in Fig. \ref{fig:Photosphere}, we find that the L band flux is in good agreement with \bpic's photosphere model prediction. The lack of infrared excess at $3.77\,\mu$m and the infrared excess at $5\,\mu$m indicates that there could be a sudden turn-on of infrared infrared excess in that region. This tentative, weak infrared excess indicates that there might be a population of host dust emitting in the wavelength range of $4$--$5\,\mu$m. We discuss the implication of this result in section \ref{dis:hotIR}. 

We find that the $3$--$5\,\mu$m excess is probably due to thermal emission from hot dust. To first order, light that is scattered off of dust grains has the same SED as the stellar host star. Our discovered  $3$--$5\,\mu m$ excess does not have a color consistent with a Rayleigh-Jeans black body as would be expected for the SED of an A-type star at $3$--$5\,\mu m$. Second, the magnitude of scattered light from the disk is significantly smaller than the observed infrared excess. Even in the brightest cases (STIS scattered light images of debris disks), only $\sim\,0.1$\% of the incident starlight is scattered by the dust \citep[e.g.,][]{Schneider+14}. At $5\,\mu m$, the IRS AdOpt spectrum has a $30\%-50\%$ excess flux with respect to the flux predicted from its stellar photosphere model. Therefore, we exclude the scattered light hypothesis.

\section{Spectral Feature Fitting}
\label{section:fitting}

In this section, our objective is to find the best-fit models of grain properties---composition, size, shape and temperature---for the \bpic~AdOpt spectrum. To obtain these properties, we first construct a disk model. Next, we select a suite of lab-measured dust optical constants and use them to calculate dust emissivities by varying grain properties. We then describe our fitting procedure. Finally, we report our best-fit models and immediate findings from these models. 

\subsection{Modeling Disk Continuum Emission}
\label{sub:continuum}
\begin{figure*}[t]
\epsscale{1}
\plotone{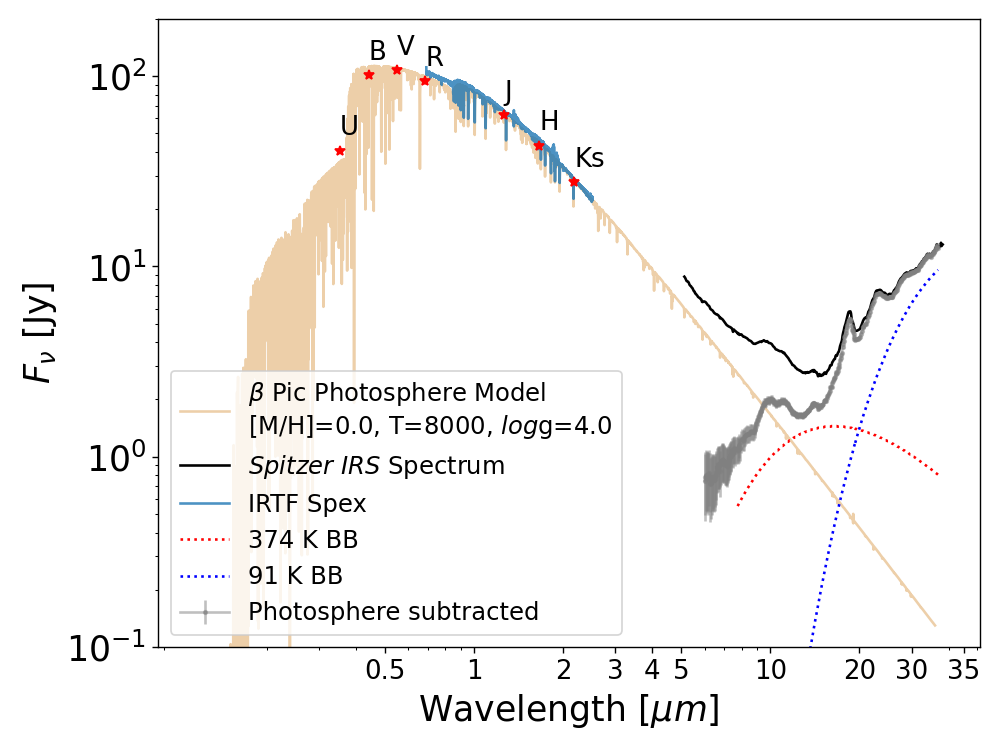}
\caption{\bpic~ SED with best fit black bodies overlaid. The cool dust component has a temperature $91$K (blue dotted line) while the warm dust component has a temperature $374$\,K (red dotted line). We use these black body temperatures to select the temperature-dependent forsterite and enstatite optical constants used in our spectral feature fitting.
}\label{fig:BB}
\end{figure*}

To isolate the solid state emission from the disk continuum emission, we first model the disk continuum emission by fitting two black body components to it. As large grains (10-100$\,\mu m$) mainly contribute to disk continuum emission, most disk continua can be modeled by a two black body model \citep{Mittal+15}. We use $5.61$--$7.94$, $13.02$--$13.50$, $14.32$--$14.83$, $30.16$--$32.19$, and $35.07$--$35.92\,\mu m$ regions as anchoring points to fit for two black bodies. We find the disk continuum consists of a warm black body at $\sim 374\pm 80\,K$ and a cool black body at $\sim 90\pm10\,K$ by minimizing the $\chi^2$ value. We plot the two black bodies alongside the disk spectrum in Figure \ref{fig:BB}. The $\sim 300\,K$ and $\sim 100\,K$ black bodies are later used for estimating the temperatures of the small grains, that are responsible for the solid state emission features. Finally, to obtain a spectrum with only solid state emission features, we subtract the 2 fitted black bodies from the IRS photosphere-subtracted \bpic~disk spectrum. In the following sections, we work with this version of the spectrum.

\subsection{Modeling Solid State Emission Features}\label{sub:solidstate_Model}
We investigate the mass and composition of the silicate grains in the \bpic~debris disk by fitting the solid-state features in the IRS spectrum. We model the emission from small grains assuming the Rayleigh limit, in which the grain sizes are much smaller than the wavelength of the incident light ( $a\ll\frac{\lambda}{2\pi}$). Such small grains are responsible for sharp, well-defined spectral features, while the large grains (in \bpic's case, $10\,\mu$m or larger) can only produce very flat spectral features \citep[e.g.,][their Fig. 6, 7, 8]{Kessler-Silacci+06}. Specifically, for \bpic, past analyses have revealed that sub-$\mu$m sub-blowout-sized grains are abundant in the \bpic~disk, indicating an active dust replenishing mechanism \citep[e.g.,][]{Okamoto+04, Czechowski+Mann07, Dent+14, Kral+16}. In addition, as these sub-$\mu$m-sized grains are expected to be the most abundant grain sizes in \bpic, they dominate the emission cross-sections according to the power-law number distribution \citep[e.g. $f(a) \propto a^{-3.5}$,][]{Dohnanyi+69, Pan+Schlichting12}{}.

To describe the size distribution of these sub-$\mu$m-sized grains, we use 3 shape distributions: spheres (Mie theory), a Continuous Distribution of Ellipsoids characterized by equal probability of all shapes (CDE1 hereafter) and by a quadratic weighting in which extreme shapes such as plates and needle have been removed (CDE2) \citep[for the analytical functions of CDE1 and CDE2, see ][]{Fabian+01}. 
The ellipsoids in the CDE distributions can range from prolate ellipsoids such as needles and footballs to oblate ellipsoids such as pancakes and plates. The CDE approximations not only offer analytical solutions for the mass absorption coefficients (MAC), but can also be computed fast enough to quickly explore a large parameter space for different dust compositions at different temperatures and abundances. 

We assume that the \bpic~debris disk is optically thin at all \textit{Spitzer} IRS wavelengths, and therefore the exact solution to the disk model is simply the sum of the emission from each grain population. Previous models have shown that the \bpic~system is well-approximated using two thin dust rings, each with distinct composition and temperature \citep[e.g.,][]{Li+Greenberg98,Chen+07}. We follow this convention and assume separate temperatures and compositions for each of the two populations.  Our model of the disk is similar to that used by \citep{Sargent+09b} to model protoplanetary disks, 
\begin{equation}
    \begin{split}
F_{\nu} & =  B_{\nu} (T_c) ( a_{c,0} + \sum_{i} a_{c,i} \kappa_{\nu,i}) \\
&+ B_{\nu} (T_w)  (a_{w,0} + \sum_{j} a_{w,j} \kappa_{\nu,j}) ,
\end{split}\label{eq:model}
\end{equation} 
where $F_{\nu}$ is the flux at each wavelength, $B_{\nu}$ is the spectral radiance of a black body as a function of temperature, $T_c$ and $T_w$ are the temperatures of the cool and warm components of the disk respectively, and $a_{c,i}$  ($a_{w,j}$) are the mass fraction in which $a_{c,i} = m_{c,i}/d^2$ ($a_{w,j} = m_{w,j}/d^2$). $m_{c,i}$ is the mass of the i$^{th}$ (j$^{th}$)  dust grain species at $T_c$ ($T_w$) and $d$ is the distance to beta Pic. $a_{c,0}$ and $a_{w,0}$ are offset values that account for the disk continuum emission from large grains. $\kappa_{\nu, i}$  and $\kappa_{\nu,j}$ are the mass absorption coefficient in cm$^{2}$ g$^{-1}$ (or emissivity for simplicity). We note that $\kappa_{\nu, i}$ for the crystalline silicate species are temperature dependent but we omit this temperature dependence in the notation. 

Our model has a total of 16 free parameters: $T_c$ and $T_w$, the mass fractions for 6 species of dust grains with temperatures $T_c$ and $T_w$ respectively, and two offset values $a_{c,0}$ and $a_{w,0}$. Table \ref{tbl-chi2models} lists the parameters. We model our spectrum by adapting code developed by \citet{Sargent+09b}. Since the \citet{Sargent+09b} study, the library of silicate optical constants has grown, expanding to include measurements with a larger number of Fe/Mg ratios  and temperatures \citep[e.g.,][]{Zeidler+15}. We tailor the set of optical constants to better fit the \bpic~debris disk, leveraging the new laboratory measurements.  We require the mass fractions to be non-negative numbers in the fitting procedure and use $\chi^2$ optimization. If the grains were black bodies with the same emissivity, then the two populations would form two concentric thin rings, each with negligible radial width. However, since our model includes $6$ different species of dust, where each species has distinct emissivities, grains with different compositions but the same temperature will not be co-located. The model can be simply understood as a disk with 12 dust rings, with 6 rings emitting at $T_c$ and other 6 emitting at $T_w$. Note that our choice of emissivities, $\kappa$, is limited to lab-measured optical constants at $100\,K$, but in reality, the cool grains can vary from $\sim80\,K$ to $\sim120\,K$ and their emissivities will change with temperature.

\subsection{Dust Emissivity}
Past analyses based on spectra and imaging data indicate that grains are predominantly composed of silicates and organics \citep[e.g.,][]{Chen+07, Ballering+16}. Since the IRS \bpic~spectrum shows prominent emission features associated with silicates, we investigate lab-measured optical constants for silicate species. The grain composition primarily affects the central wavelength location of emission features. In addition, for every grain species, four additional grain properties (crystallinity, Fe/Mg ratio, shape, and temperature) can shift the central wavelength features around. Therefore, we explore the Jena database for the most suitable optical constant measurements.

First, we select grain species based on the observed central peak wavelengths and rule out the species from visual examinations. We select olivine and pyroxene, which have characteristic features in the $10$, $18$--$20$, $23$--$25$ and $28$--$33\mu$m regions. We exclude quartz (SiO$_2$) as quartz has sharp and triangular $9\,\mu$m features that is inconsistent with our trapezoidal $10\,\mu$m feature. We also exclude carbonates because our IRS spectra do not have any $6\,\mu$m feature that resemble their characteristic features. 

Next, we divide the olivine and pyroxene into amorphous and crystalline groups. We use amorphous pyroxene (\ch{Mg_{0.7}Fe_{0.3}SiO_3}) and crystalline pyroxene, which is known as enstatite \citep{Chihara+02}. We also use amorphous olivine (\ch{MgFeSiO_{4}}) from \citet{Dorschner+95} and crystalline olivine, which is known as forsterite from \citet{Zeidler+15, Fabian+01}.\footnote{For forsterite and enstatite, we leave their exact stoichiometry for a discussion in the next paragraph.} Since amorphous grains have only two broad emission features (at $10\mu$ and $20\mu$m), their emission features are primarily affected by their size distribution. Therefore, we calculate the emissivities of the small amorphous grains using CDE2 and the emissivities of the large amorphous grains using Mie theory and grains with a $5\,\mu$m radius. 

\begin{figure}[t]
\epsscale{1.2}
\plotone{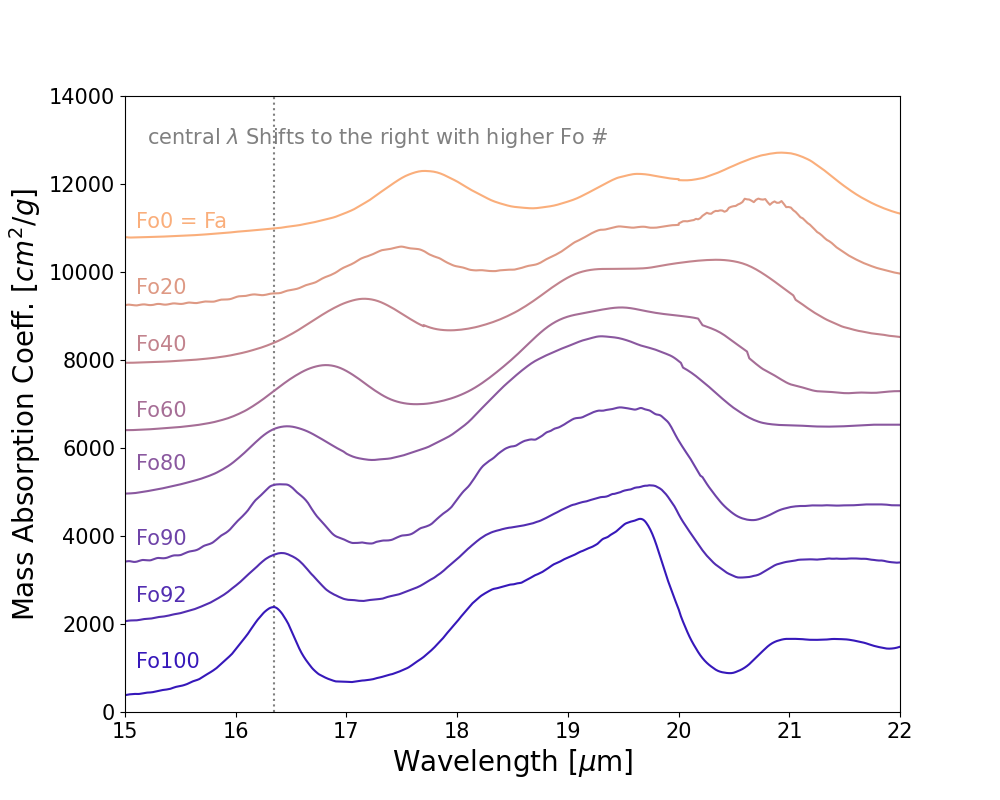}
\caption{Spectra of forsterite as a function of Fo number at room temperature ($\sim 300$ K). The silicate emission feature central wavelengths shifts towards longer wavelengths with increasing Fe abundance from bottom (Fo100 means 100\% Mg and no Fe) to the top (Fa means 100\% Fe and no Mg). To help visualize the differences among samples, we vertically offset the spectra, adding multiples of 1000 cm$^{2}$ g$^{-1}$ to each spectrum. We note that the emission features vary with both the Fo number and silicate temperature. We fix the temperature at $\sim300\,$K to showcase the effect of Fo number in this plot.}\label{fig:fo}
\end{figure}

For crystalline silicates, there are 3 grain properties that can shift the peak wavelengths of crystalline grain emission features to a shorter wavelength: (1) A decrease in Fe/Mg ratio, (2) a decrease in the grain temperature, and (3) an increase in the grain sizes or porosity. For (1), the peak wavelengths of the emission features shift toward shorter wavelengths as the Fe/Mg ratio decreases, and toward longer wavelengths as the ratio increases. We show an example in Figure \ref{fig:fo} that an $8\%$ increase in the Fe content (from Fo100 to Fo92) broadens and redshifts the bands by $0.18\,\mu$m in the $18$--$20\,\mu$m features \citep{Chihara+02,Koike+03}.\footnote{Here we use the Fo notation, where Fo stands for the percentage of Fe in forsterite stoichiometry. Fo100 represents the magnesium-rich end-member of olivine, forsterite (Mg$_{2}$SiO$_4$) with $100$\% Mg and 0\% Fe, and Fo0 represents the Fe-rich end-member of olivine, fayalite (Fe$_{2}$SiO$_{4}$, known as Fa), with $100$\% Fe and 0\% Mg. For example, Fo80 represents 80\% Mg and 20\% Fe. The same convention is used to describe the enstatite stoichiometry.} For (2), the peak wavelength of the emission feature shifts toward a longer wavelength as the temperature of the crystalline silicate increases. For example, a $200$\,K decrease in forsterite grain temperature (from $300$\,K to $100$\,K) would redshift peak wavelength by $0.09\,\mu$m from $18.95$ to $18.85\,\mu$m \citep{Zeidler+15}. According to our dust continuum model in Section \ref{sub:continuum}, we use the $300$ and  $100\,$K lab-measured forsterite and enstatite optical constants. Given that \citet{Zeidler+15} measure the optical constants of forsterite and enstatite on a sparse temperature grid of $10$, $100$, $200$, $300$, $551$, $738$ and $928$\,K, we choose not to interpolate the grid to obtain finer temperature resolutions to avoid introducing artifacts. Even though our best-fit grain temperature might deviate from these exact values by as much as $\sim80\,$K, we only use the optical constants reported in \citet{Zeidler+15}. For (3), as the grain sizes increase or become more porous, the spectral features become flatter for both amorphous and crystalline silicates, and the peak wavelength of the feature shifts towards longer wavelengths \citep{Kessler-Silacci+06}. To account for the change in grain emission feature due to size and shape effect, we create 3 different shape groups for the same species (see Table \ref{tbl-chi2models} ``Shape'' column) as aforementioned in Section \ref{sub:solidstate_Model}.

For forsterite, we experiment with the optical constants measured from the San Carlos low-Fe olivine ($98.9$\% Mg-content) sample \citep{Zeidler+15} at temperatures from 10K to 928K. We also experiment with the earlier optical constants from \citet{Fabian+01} measured at room temperature. Among all temperatures, we find that the $100\,$K, 98.9 percent Mg-rich forsterite (Fo99) provides the best wavelength match to the IRS spectrum's $18.5\,\mu$m feature. Interestingly, this Fe/Mg ratio is also consistent with the Fe/Mg ratio (99\% Mg) measured from the \textit{Herschel}/PACS $69$\,$\mu$m forsterite band \citep{deVries+12}, which is highly sensitive to the Fe/Mg ratio \citep{Koike+03}. We also include a $300\,$K Fo90 component for our warm dust component. In addition, we examine the high-Fe content cystalline olivine (known as fayalite) and find that fayalite has double emission peaks in the range of $15$--$20\,\mu$m. We exclude fayalite from our model, because its spectral features are not consistent with our observed single peak emission feature in the same wavelength range. 

For enstatite, we use temperature-dependent optical constants from \citep{Zeidler+15}. Similarly, we find that a $98.9$\% Mg-content enstatite (En99) \citep{Zeidler+15} at $300\,$K and $100\,$K have spectral features that are consistent with our IRS emission features. Therefore, we also include them in our suite of emissivities. We also find that past works \citep[e.g.,][]{Sargent+09b, Sargent+09ApJ}  demonstrated that the $95$\% Mg-content forsterite (Fo95) \citep{Fabian+01} and $90$\% Mg-content enstatite (En90) \citep{Chihara+02} produce good matches with IRS spectra. Therefore, we include an additional set of opacities (Fo95 and En90) as alternative opacities for (Fo99 and En99). 

\subsection{Fitting Procedures}
Based on our black body continuum fit described above, we constrain $T_{c}$ and $T_w$ to be within (80, 160) and (260, 380) K. Each temperature range is divided into 11 steps and hence the uncertainties for the fit are $\sigma_{T_w} = 7$\,K and  $\sigma_{T_c} = 10\,$K. We minimize $\chi^2$ for our fit by iterating over these two temperature ranges. For a more in-depth description of the fitting procedure, we refer the reader to Section 3.4 in \citet{Sargent+09ApJ} and Section 3.7 in  \citet{Sargent+09b}.

\begin{deluxetable*}{lccccc}
\tablecaption{Dust Masses for the Best-Fit Models of the \bpic~debris disk}\label{tbl-chi2models}
\tablehead{
\colhead{Species} & \colhead{Shape} &\colhead{$10\,\mu$m} &\colhead{$18\,\mu$m} & \colhead{$23\,\mu$m} & \colhead{$28$ and $33\,\mu$m}\\
\colhead{} & \colhead{} &\colhead{($10^{-3}$ M$_{moon}$)} &\colhead{($10^{-3}$ M$_{moon}$)} &\colhead{($10^{-3}$ M$_{moon}$)}&\colhead{($10^{-3}$ M$_{moon}$)}}
\startdata
\multicolumn{2}{l}{Cool Dust Continuum Temperature (T$_{c}$ )}                & $91$\,K             & $84$ \,K          & $82$\,K         & $82$\,K\\
\hline
Pyroxene (Mg$_{0.7}$Fe$_{0.3}$SiO$_{3}$)       &CDE2, Rayleigh Limit                  & $11.6\pm 2.0$         & $0\pm 2.8$      & $0\pm 3.6$     & $0\pm 3.5$\\
Pyroxene      &Mie, $5\,\mu m$ radius, $60$\% porosity & $0\pm 1.5$         & $0\pm 2.2$      & $0\pm 2.7$     & $0\pm 2.7$\\
Olivine (MgFeSiO$_{4}$)        &CDE2, Rayleigh Limit                  & $10.9\pm 1.5$      & $34.5\pm 2.3$   & $26.0\pm 2.7$  & $19.6\pm 2.6$\\
Olivine       &Mie, $5\,\mu m$ radius, $60$\% porosity & $0\pm 1.0$.        & $0\pm 1.4$.     & $0\pm 1.8$     & $0\pm 1.7$ \\
Forsterite (Mg$_{1.72}$Fe$_{0.21}$SiO$_{4}$)                     & (1)                  & $3.5\pm 0.8$      & $2514\pm 507$   & $4483 \pm 594$ & $3031\pm 425$ \\
Enstatite (Mg$_{0.92}$Fe$_{0.09}$SiO$_{3}$)                    & (1)                  & $0\pm 0.7$       & $183\pm 564$    & $260\pm 250$   & $391 \pm 474$  \\
\hline
\multicolumn{2}{l}{Warm Dust Continuum Temperature (T$_{w}$)}                & $374$\,K             & $298$\,K         & $282$\,K        & $272$\,K\\
\hline
Pyroxene (Mg$_{0.7}$Fe$_{0.3}$SiO$_{3}$)         &CDE2, Rayleigh Limit                   & $0.019\pm 0.003$  & $0.046\pm 0.006$  & $0.064 \pm 0.007$ & $0.067 \pm 0.008$\\
Pyroxene       &Mie, $5\,\mu m$ radius, $60$\% porosity  & $0 \pm 0.003$ & $0.017 \pm 0.006$ & $0.031\pm 0.008$  & $0.074\pm 0.009$\\
Olivine (MgFeSiO$_{4}$)         &CDE2, Rayleigh Limit                   & $0\pm 0.002$      & $0\pm 0.004$      & $0\pm 0.005$      &  $0\pm 0.005$\\
Olivine        &Mie, $5\,\mu m$ radius, $60$\% porosity  & $0\pm 0.003$      & $0\pm 0.005$      & $0\pm 0.006$      & $0\pm 0.007$ \\
Forsterite (Mg$_{1.72}$Fe$_{0.21}$SiO$_{4}$)                      & (1)                   & $0.0024\pm 0.0017$    & $0.82\pm 0.81$    & $0.82 \pm 1.33$   & $ 0.37\pm 1.46$ \\
Enstatite  (Mg$_{0.92}$Fe$_{0.09}$SiO$_{3}$)      & (1)                   & $0\pm 0.0016$        & $0\pm 0.8$        & $2.3\pm 1.6 $     & $1.94\pm 1.69$\\
\hline 
$\chi^2$                       & \nodata                & $23.8$            & $23.6$            & $25.3$            & $30.3$
\enddata
\tablecomments{1. The shape distribution for fosrterite and enstatite are CDE1 for$10\,\mu$m feature, Mie for $18\,\mu$m feature, CDE2 for $23\,\mu$m, CDE1 for $28$ and $33\,\mu$m features.  2. The warm dust species share the same stoichiometry as the cold dust species. 3. The optical constants for the forsterite and enstatite used for the cool dust are measured at $100$~K, while those for the warm dust are measured at $300$ K. Therefore, their mass fraction coefficients are independent from each other. 4. The $\chi^2$ values calculated for the best-fit models use the full wavelength range ($5$--$35 \mu$m) in the IRS spectrum. }
\end{deluxetable*}

\begin{figure*}[t]
\plottwo{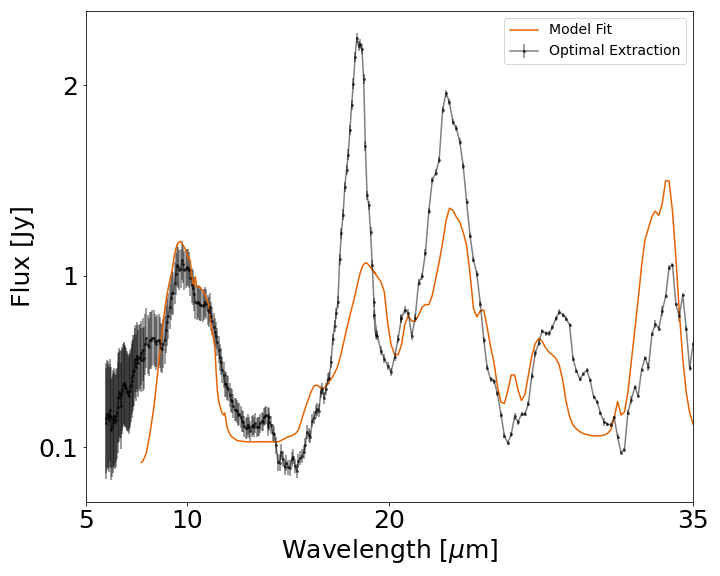}{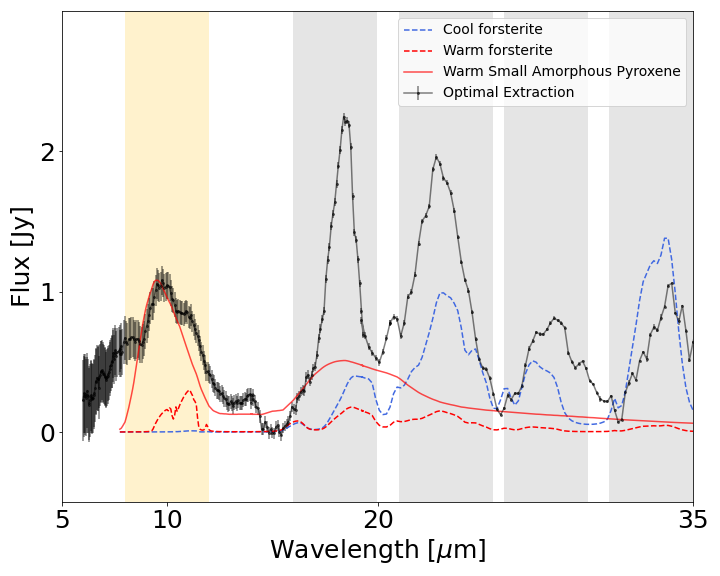}
\caption{Best-fit model for the $10\,\mu m$ region. Left: Model fit (orange) plotted over the data (black). Right: Components of different grains species plotted over the data. Warm and cool dust species are plotted in red and blue respectively. Flux from all dust species sums to the model fit (orange) in the left panel.   We use gray bands to mark the $18$, $23$, $28$ and $33\,\mu$m bands where the model does not fit the data well and use pale yellow bands to highlight the areas where the model fits the data well.  
\label{fit10mu}}
\end{figure*}

\begin{figure*}[t]
\plottwo{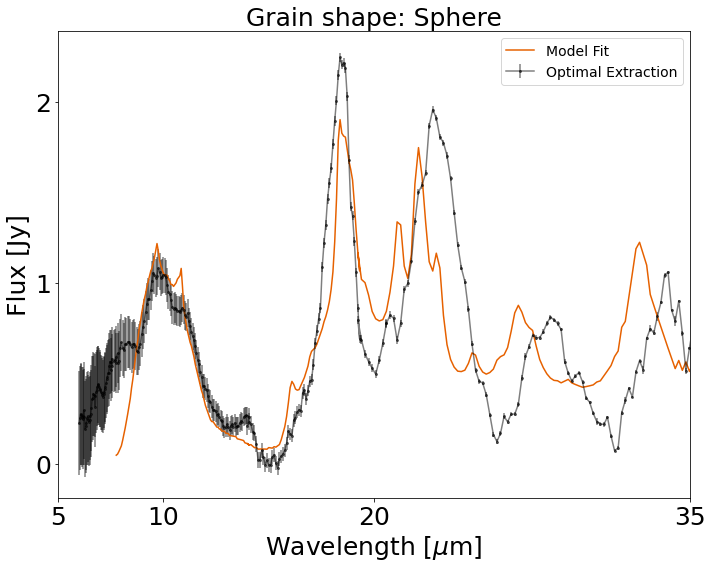}{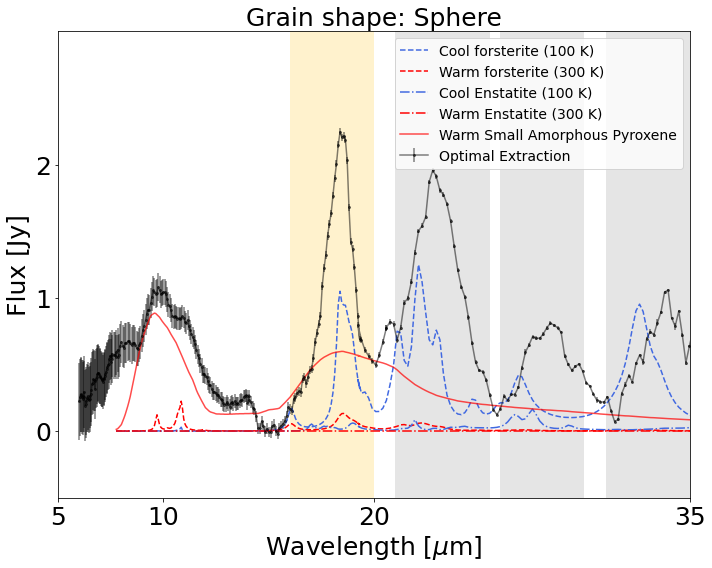}
\caption{Best-fit model with particle shape calculated with the Mie thoery. We show the disk spectrum with its featureless black bodies components subtracted from the total flux. Left: Model fit (orange) plotted over the data (black). Right:Components of different grain species plotted over the data. Warm and cool dust species are plotted in red and blue respectively. Flux from all dust species sums to the model fit (orange) in the left panel. We use gray bands to mark the $23$, $28$ and $33\,\mu$m bands where the model misfits and use pale yellow bands to showcase where the model performs well.}  
\label{fitSphere}
\end{figure*}

\begin{figure*}[t]
\plottwo{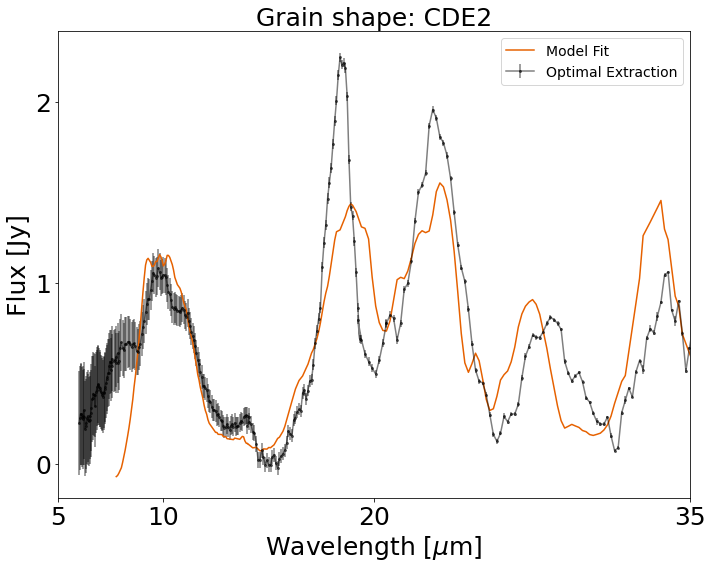}{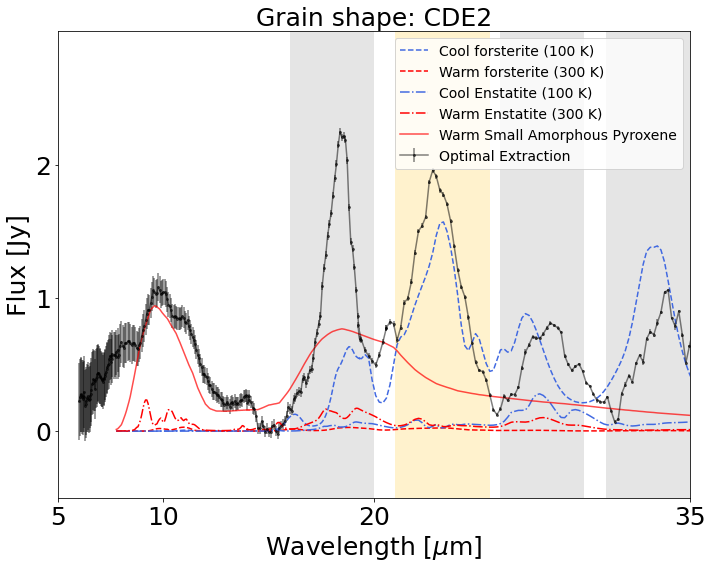}
\caption{The Best-fit model with particle shape calculated with the continuous distribution of ellipsoids \citep[CDE2,][]{Fabian+01} with quadratic weighting for forsterite grains and enstatite grains. We show the disk spectrum with its featureless black bodies components subtracted from the total flux.  Left: Model fit (orange) plotted over the data (black). Right: Components of different grain species plotted over the data. The flux from all the grain species (blue and red lines) sums up to the total flux, which equals to the model fit (orange) in the left panel.  \label{fitCDE2}}
\end{figure*}

\begin{figure*}
\plottwo{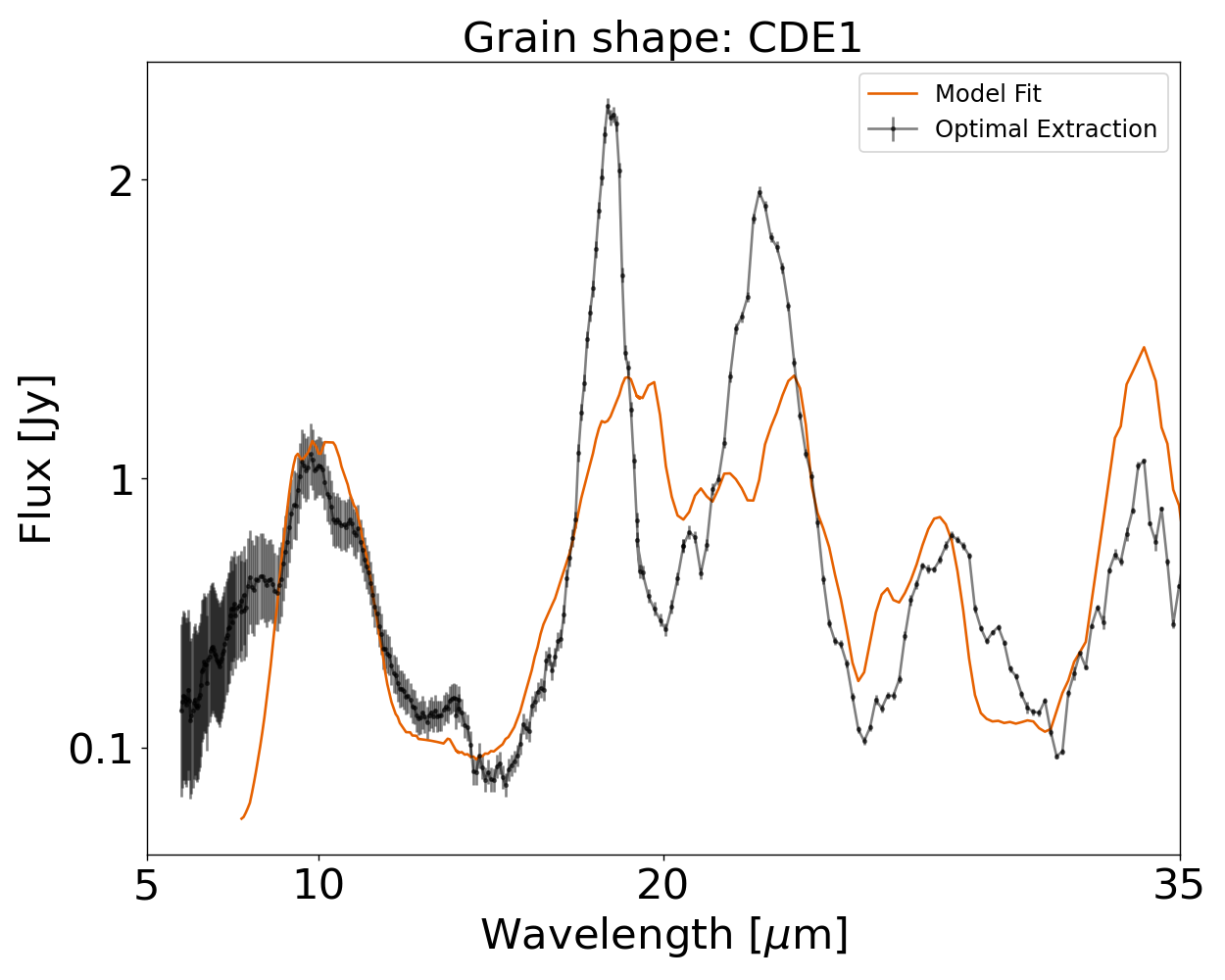}{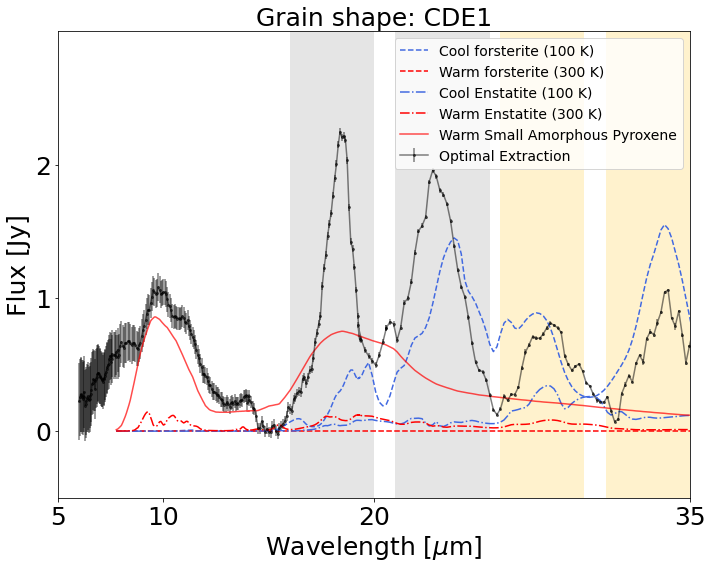}
\caption{The Best-fit model with particle shape calculated with the continuous distribution of ellipsoids \citep[CDE1,][]{Fabian+01} for forsterite grains and enstatite grains by assuming all shapes are equally probable. We show the disk spectrum with its featureless black bodies components subtracted from the total flux. Left: Model fit (orange) plotted over the data (black). Right: Components of different grain species plotted over the data. Warm and cool dust species are plotted in red and blue respectively. Flux from all dust species sums to the model fit (orange) in the left panel \label{fitCDE1}}
\end{figure*}

\subsection{Best-Fit Models}
We find that \bpic~at minimum contains a warm ($\sim300\,$K) and a cool ($\sim100\,$K) population of dust, where the cool population is mainly responsible for the $18$--$33\,\mu$m emission. We plot our best-fit models in Figure \ref{fit10mu}, \ref{fitSphere}, \ref{fitCDE2}, and \ref{fitCDE1} and tabulate their various species of dust masses in Table \ref{tbl-chi2models}. The best-fit model shows that the $18\,\mu$m feature is mostly emitted by the $100$\,K cool forsterite. The best-fit model contains $3000$ times more cool forsterite mass than the $300$\,K warm forsterite mass. 

Even though the $18\,\mu$m feature is well modeled in Figure \ref{fitSphere}, the $23$, $28$ and $33\,\mu$m features are not well modeled. The model-predicted features all have peak wavelengths shorter than the observed peak wavelengths. In Figure \ref{fitSphere}, gray bands mark the $23$, $28$ and $33\,\mu$m bands where the model does not fit the data well and pale yellow bands showcase where the model performs well.

By modifying our grain shapes to be a moderate continuous distribution of ellipsoids (CDE2), we optimize the fit of the model for the $23\,\mu$m spectral feature. Similarly, by modifying our grain shapes to contain extreme shapes in a continuous distribution of ellipsoids (CDE1), we optimize the fit of the model for the $28$ and $33\,\mu$m spectral features. However, any particular shape distribution (whether spherical, CDE2, or CDE1) only improves the model fit for a localized region ($18$, $23$ or $28$\&$33\,\mu$m) in the spectrum. We experiment with a 4-grain model with 4 different temperature components, but find that increasing the number of model parameters does not improve the quality of the fit. The 4-grain model produces a similar $\chi^2$  value ($\chi^2\approx26$) to the 2-grain models. Therefore, we report numbers for the 2-grain models. In addition, we also experiment with fitting only an isolated region in the spectrum (instead of the entire spectrum). For example, we fit the $18\,\mu$m feature by minimizing only the residual between the model and the data in the $15$--$20\,\mu$m region. However, the model wildly overpredicts the flux in the $10\,\mu$m region by more than $200\%$, sacrificing all other spectral features to optimize one single feature. These undesirable results with alternative models motivate our choice to minimize the residual over the entire wavelength range.

From the dust masses reported in Table \ref{tbl-chi2models}, we find that the dust population responsible for the $18$--$33\,\mu$m features consists of more than $90$\% sub-$\mu$m-sized crystalline grains and less than $10$\% of sub-$\mu$m-sized amorphous grains in mass. Our best-fit models indicate that the amorphous pyroxene and olivine grains with radii larger than $5\,\mu$m cannot account for the solid state emission features in the IRS spectrum. Specifically, in Table \ref{tbl-chi2models}, the coefficients for ``pyroxene (Mie, 5 $\mu$m radius, 60\% porosity)'' and ``olivine (Mie, 5 $\mu$m radius, 60\% porosity)'' are all consistent with 0.

Furthermore, for cool dust population, we find that the sub-$\mu$m grain shape becomes increasingly irregular with increasing wavelength. For the terrestrial-temperature, warm dust, the $10\,\mu$m feature is best-fit using CDE1 grain shapes, which indicates that the grain shapes are irregular. For the cool dust grains, the best-fit models require different grain shape distributions for $18$, $23$ and $28$\&$33\,\mu$m features to optimize the model's $\chi^2$ value. The $18\,\mu$m feature is very sharp and is best fitted using spherical grains, while the $23\,\mu$m feature is best-fit using CDE2, and the $28$ and $33\,\mu$m features are best-fit using CDE1.

To conclude, we find that the grain's properties, such as shape, crystallinity, and composition, change with increasing wavelength. In the next section, we examine if there is a trend in silicate crystallinity and Mg/Fe abundance as a function of stellocentric distance, of which wavelength is a proxy.

\section{Abundance Analysis} \label{section:abundance}
In this section, we investigate whether there is a trend in (1) crystallinity and (2) Fe/Mg in small grains as a function of wavelength, a proxy for stellocentric distance, in the \bpic~debris disk. The crystallinity and Fe/Mg ratio inform us about the formation conditions and origins of these silicate dust grains. In section \ref{section:fitting}, we discover that sub-$\mu$m-sized grains are responsible for all prominent $10$--$33\,\mu$m features, where each feature is best fitted by a separate population of grains consisting of both crystalline and amorphous silicates with a distinct mass ratio and a preferential shape distribution. We group the features by their optimal fits and investigate the properties for each group as a way to probe their parent bodies' properties.

\subsection{A Crystallinity Gradient} \label{subsec:cry}
We investigate crystallinity fractions in small, sub-micron-sized grains as a function of wavelength, as a proxy for radial distances. Note that we do not investigate the  crystallinity fraction in large grains with radius equal to or larger than $5\,\mu m$, because both amorphous and crystalline grains in this size regime produce very broad and difficult to fit features.
 
Our model assumes a simplified scenario of grains emitting at two temperatures. If the grains are perfect black bodies from a single population of dust species, then this scenario can be viewed as two concentric, infinitesimally narrow rings. However, our fitting results show that the cold outer belt contains multiple populations of cool forsterite and amorphous silicates that on average emit at $\sim100\,$K. Hence, this belt must have a non-negligible radial width. To separate the different forsterite populations, we assume the silicates in the disk are in radiative equilibrium and consider grains as black bodies. Then we calculate the black body temperatures that correspond to the peak wavelength of the emission feature with Wien's law and obtain a black body distance. The $10$, $18$, $23$, $28$, and $33\,\mu$m silicate emission features correspond to black body temperatures of $290$, $160$, $126$, $103$, and $83\,$ K and radial distances of $3$, $9$, $14$, $21$ and $33\,$AU in the disk, respectively. Note that these distances are lower limits for the actual radial distances, because small grains have lower emission efficiency than black bodies and can stay warm at further radial distances. Note also that our 2-temperature model cannot accurately constrain the temperatures of different cool forsterite populations, as our data is limited to the $100K$ forsterite opacity.

\begin{figure}
    \epsscale{1.2}
    \plotone{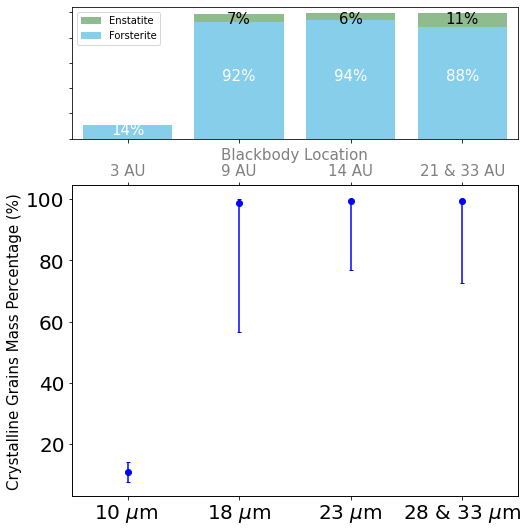}
    \caption{The crystallinity of small silicate grains as a function of radial distance as represented by spectral features emitting at increasing larger disk radii. The y axis shows the percentage of different grain compositions by mass. The x axis shows the best-fit models for $10$, $18$, $23$, $28$ and $33$ $\,\mu$m silicate features.  We calculate a black body distance at each wavelength by assuming that the grains are black bodies and in thermal equilibrium such that the incident stellar radiation on a grain equals to its thermal re-emission. Note that these distances are lower limits to the actual distances where the grains reside because the grain opacity change with wavelength, where the black body's opacity do not have a wavelength dependence.}
    \label{fig:cry}
\end{figure}

We use the values in best-fit models at $10$, $18$, $23$, $28$ and $33$\,$\mu$m silicate emission features as presented in Fig. \ref{fit10mu}, \ref{fitSphere}, \ref{fitCDE2} and \ref{fitCDE1} to calculate the abundance of the four silicate species. The crystalline silicate species are enstatite (Mg$_{0.92}$Fe$_{0.09}$SiO$_{3}$) and forsterite (Mg$_{1.72}$Fe$_{0.2}$SiO$_{4}$), and the amorphous silicate species are olivine and pyroxene. In Fig \ref{fig:cry}, we calculate the mass percentages of enstatite (green), forsterite (blue), and amorphous silicates (gray) as a function of radial distance. 

The crystalline fraction in small grains increases from $14\pm3$\% at $10\,\mu$m ($\sim$3~AU) to $99^{+1}_{-42}$\% at $18\mu$m  ($\sim$10~AU) and remains high from  $18\mu$m  ($\sim$10~AU) to $33\mu$m ($\sim$33~AU). This sudden increase in grain crystallinity at $\sim$10~AU highlights a major change in the crystallinity of the grain composition. We tabulate the crystallinity fractions in Table  \ref{tbl-femg-crystallinity}. Applying these trends to \bpic~planetary architecture, we find that the sub-micron-sized silicate grains exterior to the \bpic~b ($10$~AU) are highly crystallized, while the ones interior to \bpic~b are mostly amorphous.

To understand the abundance of crystalline grain in the \bpic~debris disk, we examine their production mechanisms. Two processes produce crystalline grains in a debris disk: (1) Thermal annealing and (2) collisional grinding between parent bodies with crystalline silicate-rich surfaces. Thermal annealing is a process in which amorphous silicates are heated to high temperatures (but below their vaporization temperatures) for enough time that their internal structure rearranges to be crystal-like, producing forsterite, enstatite, and silica \citep{Henning2010}. Since thermal annealing is more likely to occur closer to the star, the abundance of crystalline silicates should be the highest closer to the star and decrease with increasing radial distance. However, we find the opposite trend in crystallinity to that predicted by the thermal annealing scenario, which rules out thermal annealing as the main crystallization mechanism. Alternatively, if the surfaces of the parent bodies are crystalline-rich, continuous collisions among these parent bodies can graze off the surface materials and produce grains with high crystallinity. In our solar system, the A-type asteroids in the main asteroid belt are known to have olivine-dominated surfaces based on their reflection spectra \citet{demeo+2019}. In debris disks, collisional grinding of crystalline-rich parent bodies' surfaces are also thought to generate enstatite-rich dust grains \citep{Fujiwara+10, Olofsson+09}.  Therefore, the collisional grinding production of crystalline silicates remains a possible production mechanism to explain the crystallinity trend in \bpic. 

\subsection{The Fe/Mg Abundance Ratio as a function of distance}
\begin{figure}
    \epsscale{1.2}
    \plotone{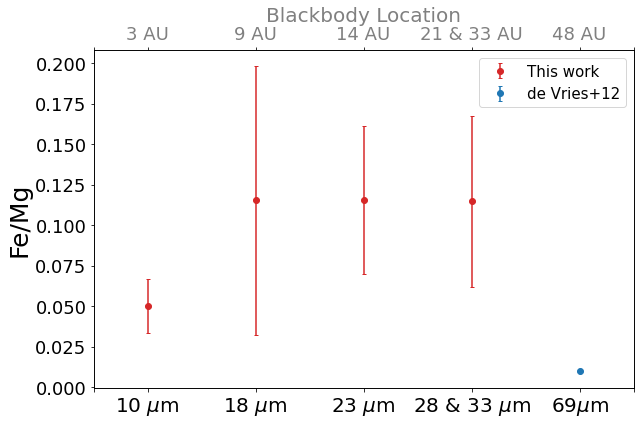}
    \caption{The Fe to Mg ratio for small grains plotted as a function of spectral feature wavelength that can be used as a proxy for stellocentric distancec. The red dots with error bars are from this work and the blue dot is the Fe to Mg ratio measured from Herschel PACS spectra \citep{deVries+12}. }
    \label{fig:ratio}
\end{figure}

To constrain the formation conditions of the parent bodies in the debris disk, we investigate the Fe/Mg ratio in the crystalline silicates in the disk. We calculate the Fe/Mg ratio by using the silicate masses reported in Table \ref{tbl-chi2models} and converting the reported masses into molecular abundance in moles. We then calculate the absolute abundances of Fe and Mg from the stoichiometry of the chemical formulas reported in Table \ref{tbl-chi2models}. We report the Fe/Mg ratio in Table \ref{tbl-femg-crystallinity} and plot the Fe/Mg ratio in crystalline silicates as a function of radial distance in Figure \ref{fig:ratio}.

The Fe/Mg ratio remains constant from $10$ to $33\,\mu$m, but decreases to less than $1$\% at $69\,\mu$m, when we incorporate the Fe/Mg abundance of $(1\pm 0.1)\%$ measured from \textit{Herschel}/PACS $69\,\mu$m forsterite feature \citep{deVries+12}. Applying this trend to the \bpic~planetary architecture, we find that the small dust grains interior to \bpic~b ($10$~AU) are more Fe-rich while the dust grains exterior to \bpic~b becomes increasingly Fe-poor. We further discuss the implications of this trend on parent body properties in section \ref{sub:surface} and \ref{sub:mineralogy}. 

In addition, we compare the \bpic's Fe/Mg ratio with the Fe/Mg ratio measured from white dwarf atmosphere compositions. Recent measurements of precise elemental abundances from white dwarf atmospheres enables us to probe the compositions of extrasolar rocky planetesimals. Interesting, we find that our reported Fe/Mg olivine ratio for the $10\,\mu$m warm dust is consistent with that of G29-38 \citep{Xu+14}, suggesting that G29-38's rocky planetesimals could contain olivine. We also tabulate the Si, O, Mg, and Fe abundances as a function of radial distance (see Table \ref{tbl-abundance-elements}). We find that the abundance of all $4$ species in small grains increases by a factor of $100$ from $22$ log(Mole) at $10\,\mu$m ($\sim300$\,K) to $24$ log(Mole) at $18\,\mu$m ($\sim160\,$K).

\begin{deluxetable*}{lcccc}
\tablecaption{Fe/Mg Ratio for Crystalline Silicates and Crystallinity Fraction for Grains as a Function of Wavelength}\label{tbl-femg-crystallinity}
\tablehead{
\colhead{Quantity} &\colhead{$10\,\mu$m} &\colhead{$18\,\mu$m} & \colhead{$23\,\mu$m} & \colhead{$28$ and $33\,\mu$m}}
\startdata
Fe/Mg & $ 0.05 \pm 0.017 $ & $ 0.12 \pm 0.08 $ &  $ 0.12 \pm 0.05 $ & $ 0.11 \pm 0.5 $ \\ 
Crystallinity (\%) & $ 14 \pm 3 $ & $ 99^{+1}_{-42} $ & $ 99.5^{+0.5}_{-23} $  & $ 99.5^{+0.5}_{-27} $ \\ 
\enddata
\end{deluxetable*}

\begin{deluxetable*}{lcccc}
\tablecaption{Element Abundance for the Best-Fit Models of the \bpic~debris disk}\label{tbl-abundance-elements}
\tablehead{
\colhead{Species} &\colhead{$10\,\mu$m} &\colhead{$18\,\mu$m} & \colhead{$23\,\mu$m} & \colhead{$28$ and $33\,\mu$m}\\
\colhead{} & \colhead{[$\log$(Mole)]} &\colhead{[$\log$(Mole)]} &\colhead{[$\log$(Mole)]}&\colhead{[$\log$(Mole)]}}
\startdata
Si & $ 22.2 \pm 0.7 $ & $ 24.3 \pm 0.2 $ &  $ 24.5 \pm 0.4 $ & $ 24.4 \pm 0.4 $ \\ 
O & $ 22.8 \pm 0.7 $ & $ 24.8 \pm 0.2 $ &  $ 25.1 \pm 0.5 $ & $ 24.9 \pm 0.4 $ \\ 
Mg & $ 22.2 \pm 0.7 $ & $ 24.5 \pm 0.3 $ &  $ 24.7 \pm 0.6 $ & $ 24.6 \pm 0.5 $ \\ 
Fe & $ 21.9 \pm 0.8 $ & $ 23.5 \pm 0.3 $ &  $ 23.8 \pm 0.6 $ & $ 23.6 \pm 0.5 $ \\
\enddata
\end{deluxetable*}

\section{Discussion}
\label{section:discussion}

\subsection{Trends in Silicates and Implications on Parent Body Surface Properties}
\label{sub:surface}

In section \ref{section:abundance}, we report our findings that the sub-$\mu$m-sized silicate grains are increasingly crystalline, Mg-rich (Fe-poor), and irregular in shape as stellocentric distance increases. We highlight that this critical transition in silicate properties occurs in the vicinity of \bpic~b's orbit. As sub-$\mu$m-sized grains must be constantly replenished from planetesimal collisions on orbital timescales, short compared to the age of the disk,  these grains reflect the surface composition and conditions of their parent planetesimals. 

From the Fe/Mg trend in silicates, we infer that the surfaces of planetesimals interior to \bpic~b are more Fe-rich compared to the surfaces of planetesimals exterior to \bpic~b. As Fe-bearing silicates are preferentially produced by planetesimal collisions, as we argue in section \ref{section:abundance}, the planetesimals close to or interior to \bpic~b might have experienced more collisions compared to the planetesimals exterior to \bpic~b (outward of $10\,$AU). If the parent planetesimals have not fully differentiated, then the bulk composition of the planetesimals could be increasingly Fe-poor as stellocentric distance increases.

From the crystallinity trend in silicates, we infer that the surfaces of planetesimals interior to \bpic~c ($\sim3\,$AU) are mostly amorphous, while the surfaces of planetesimals exterior to \bpic~b are highly crystalline. As crystalline olivine can be easily turned into amorphous olivine via collisions \citep{Henning2010}, the highly crystallized silicate surfaces of planetesimals exterior to \bpic~b indicate that these planetesimals have not undergone major collisions.

\subsection{Comparing mineralogy of \bpic~and Solar System}\label{sub:mineralogy}

We compare the Fe/Mg gradient of the \bpic~chemical reservoir with that observed in the Solar System. In the Solar System, comets that originate from the Trans Neptunian region contain Mg-rich silicates \citep{Wooden+17}, whereas asteroids and chondrites are Fe-rich \citep{LeGuillou+15}.  Such an Fe/Mg trend in the Solar System also leaves imprints on terrestrial planetary surfaces. Specifically, most surfaces on Mars contain Fo68, an olivine that is relatively Fe-rich, while ancient craters on Mars contains Fo91, which is relatively Fe-poor; Fo91 is thought to originate from the Kuiper belt \citep{Hamilton10}. Therefore, \bpic~ and the Solar System share a similar Fe/Mg trend. 

We also compare the olivine grain shape distributions in \bpic~with the olivine grain sizes observed in our Solar System. In the Solar System, sub-$\mu$m-sized forsterite grains are abundant in the chondritic porous interplanetary dust particles (IDP) and comets that originate from the Kuiper belt \citep{Wozniakiewicz+12}, while much larger and less porous forsterite grains are present in asteroids. In contrast, in the \bpic~system, we see sub-$\mu$m forsterite grains throughout the entire disk; their shapes become increasingly irregular as a function of radial distance. Simulations have shown that the fluffy, irregular grain aggregates can produce qualitatively similar spectral features as porous grains \citep{Kolokolova+10}. If we consider grain shape a proxy for grain porosity, then the sub-$\mu$m-sized, irregular forsterite grains in the outskirt of the \bpic~disk correspond to the forsterite grains in our Solar System's comets and IDP.

Our discovery of these similarities paves the way for future space-based spatially-resolved MIR spectroscopy studies.
Higher spatial resolution observation of JWST will improve our understanding of \bpic's planetary system architecture. Future JWST GTO (ID: $1294$) observations will be able to probe the spatial distribution of the the forsterite population with an improved resolution (by a factor of $\sim 10$ in spatial resolution compared to Spitzer's sptial resolution) with the MIRI Medium Resolution Spectroscopy as well as a higher overall SNR (compared to ground-based high-resolution observations).   The Gemini T-ReCS images revealed an asymmetric dust distribution at $18.3\,\mu$m in the disk \citep{Telesco+05}, while the \textit{Spitzer} discovery provides complimentary spectroscopic information by pinpointing the $18 \,\mu$m forsterite emission which give good constraints on grain properties such as Fe/Mg and crystallinity and shape. JWST would be able to map and compare the size, shape and mass distributions of cool forsterite grains in the southwest and  northeast side of the disk, leveraging the knowledge from Gemini and \textit{Spitzer}. The future JWST MRS observations will also potentially resolve more solid state emission features and therefore refine measurements of Fe/Mg and crystallinity ratio for the $18$--$33\,\mu$m features.

\subsection{Tentative Evidence for Weak $3$--$5$\,$\mu$m Hot Dust}\label{dis:hotIR}

In section \ref{sub:hotdust}, we show that a population of $\sim600$\,K hot dust population located within $0.7\,$AU likely contributes to $\sim50$\% of excess flux at around $3$--$5\,\mu$m. In comparison, past H and K band interferometric measurements indicate a $\sim1500$\,K hot dust population located at least within $4\,$AU to the star \citep{Ertel+14, Defrere+12}. It is uncertainty whether the tentative $\sim600$\,K hot dust is related to the  $\sim1500$\,K hot dust population. The $\sim600$\,K hot dust could have multiple origins. ALMA dust continuum images \citep[e.g.,][]{Kral+16} have shown that the inner disk does not have obvious cavity and can still contain abundant small planestesimals to produce the $\sim600$\,K hot dust. Alternatively, inward P-R drag could operate to move the $10\,\mu$m warm dust grains into inner region of the disk. It is also possible that there are stochastic events in the inner region of the disk such as comets infalling activities \citep{Kiefer+14} could produce the $\sim600$\,K hot dust population. Future observations are needed to confirm the tentative $\sim600$\,K hot dust population. \bpic~is too bright to be observed by \textit{JWST} from space but can be observed from the ground. IRTF is a suitable facility for obtaining the $2.5$--$5\,\mu$m NIR spectrum with high SNR to constrain the level of hot dust excess in this system.

\section{Conclusion}
\label{section:conclusion}

We re-analyze the \textit{Spitzer} IRS data of the \bpic~debris disk with AdOpt \citep{Lebouteiller+10}. To better constrain the stellar parameters, we obtain a new NASA IRTF SpeX spectrum from $0.7$--$3\,\mu$m and found a weak $3$--$5\,\mu$m excess possibly due to hot dust close to the star. We discover a prominent $18\,\mu$m silicate feature for the first time and an enhanced $23\,\mu$m feature. These narrow spectroscopic features placed good constraints on grain properties such as Fe/Mg ratio, crystallinity and shape. We find that the $\sim100$\,K forsterite grains are the main contributors to these two features. Furthermore, we find three trends in grain properties as functions of wavelengths: (1) The Fe/Mg ratio in silicates decreases with stellocentric distances. We infer that the surface composition of the planetesimals is increasingly Mg-rich and pristine the further away they are from the star. We find \bpic's chemical gradient offers an analogy to our Solar System's clearly divided grain chemical reservoirs; (2) the grains become more crystalline with increasing wavelengths and (3) lastly, the grain shapes become increasing irregular with increasing wavelengths. The findings imply that the properties of dust population in the vicinity of \bpic~b and c differs significantly in crystallinity, shape and Fe/Mg ratio. This is the first time that such a trend in spectral features has been studied with space-based MIR spectroscopy for a debris disk. Future \textit{JWST} MIRI observations will constrain the spatial location of the grains that are responsible for the newly discovered $18$ and $23\,\mu$m spectral features and probe \bpic~b's atmospheric cloud composition for comparison with dust properties in the planet's vicinity.

\begin{acknowledgements}
We thank  Alycia Weinberger, Benjamin Montesinos, Francisca Kemper, John Debes, Roc Cutri, Bin Ren and Jingwen Fan for useful discussions. CL acknowledges support from STScI Director's Research Fund (DRF) and NASA FINESST grant.
This work is supported by the National Aeronautics and Space Administration under Grant No. 80NSSC21K1844 issued through the Mission Directorate.
This work is based on archival data obtained with the Spitzer Space Telescope, which was operated by the Jet Propulsion Laboratory, California Institute of Technology under a contract with NASA. 
This research has made use of the NASA/IPAC Infrared Science Archive, which is funded by the National Aeronautics and Space Administration and operated by the California Institute of Technology. This publication makes use of data products from the Wide-field Infrared Survey Explorer, which is a joint project of the University of California, Los Angeles, and the Jet Propulsion Laboratory/California Institute of Technology, and NEOWISE, which is a project of the Jet Propulsion Laboratory/California Institute of Technology. WISE and NEOWISE are funded by the National Aeronautics and Space Administration. This research has made use of NASA's Astrophysics Data System.  This research has made use of the SIMBAD database, operated at CDS, Strasbourg, France \citep{Wenger+2000}.  This research has made use of the VizieR catalogue access tool, CDS, Strasbourg, France (DOI: 10.26093/cds/vizier). The original description of the VizieR service was published in 2000, A\&AS 143, 23 \citep{Ochsenbein+2000}.
\end{acknowledgements}

\clearpage
\begin{appendix}
\section{Spitzer IRS Spectrograph Slit Orientation}
\begin{figure}[h!]
\epsscale{1}
\plottwo{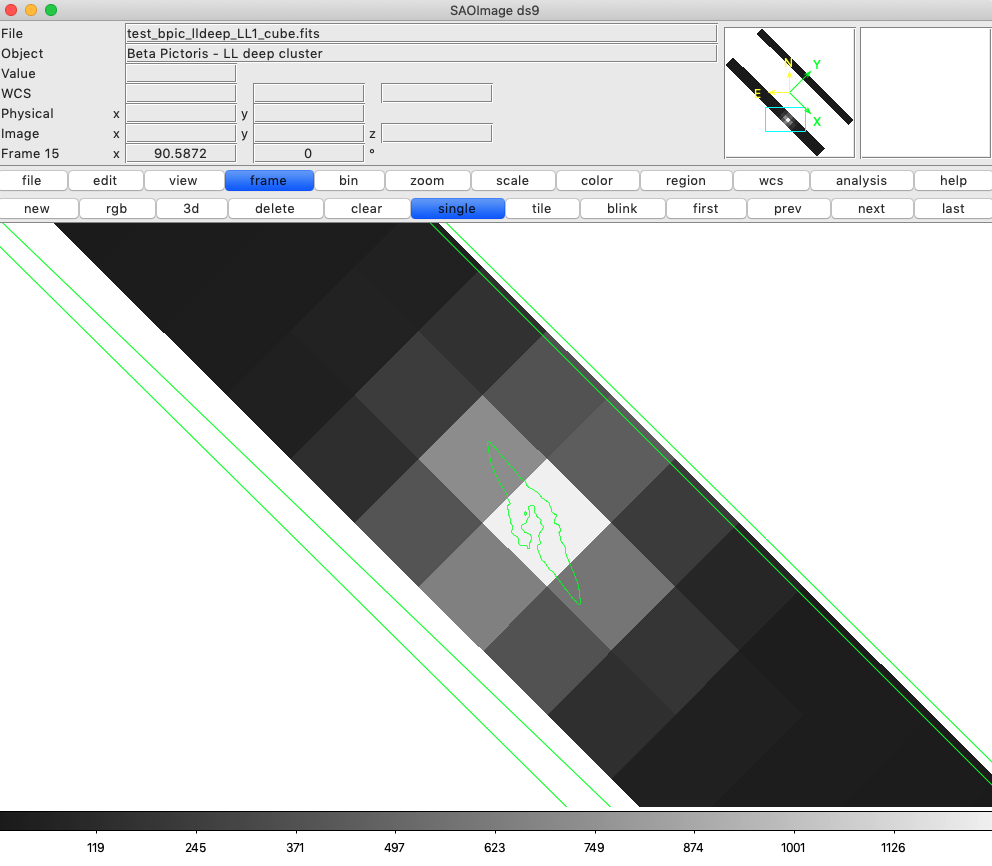}{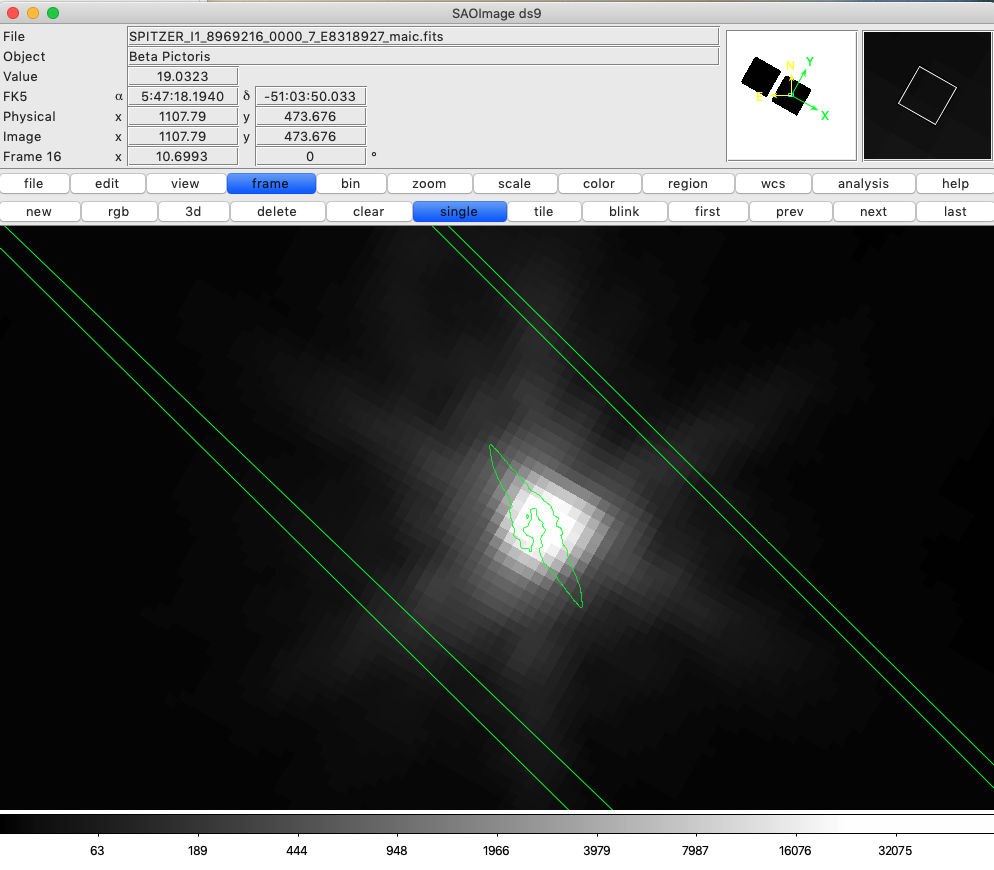} 
    \caption{The \bpic~debris disk in the slit of IRS spectrograph. The coordinate display here is the world coordinate system (WCS), in which North points upwards and East points to the left. The green contour shows the disk of \bpic~in scattered light, and is obtained from the scattered light image with the Space Telescope Imaging Spectrograph (STIS) on the Hubble Space Telescope (HST) \citep[][See Figure 6c therein.]{Ren+17}. The South-West side of the disk constains a gas clumps as seen in ALMA. The disk is measured to have a position angle of $\text{PA}_{\text{disk}}$ = $29.51^\circ$. The edges of the IRS spectrograph slit indicated by the green lines (we zoom in to show part of the slit, since the entire slit is too long). The slit has a position angle of $\text{PA}_{\text{slit}}$ = $44.97 \pm 0.02^\circ$. The disk is $15^\circ$ misaligned with respect to the slit. Left: The mosaic image in the background is the $\beta$ Pic 2D spectral map taken with IRS Long-low 1 (LL1) mode and assembled using \texttt{Cubism} \citep{IRScubism}. Shown here is the disk at $20.52~\mu m$, and there are 100 other slices (not shown here) in LL1 that span $\lambda \sim20~\mu m$ to $39~\mu m$ in wavelength.The disk roughly spans $2.5$ pixels in this image and each pixel is square with $5.1\arcsec$ on each side. Right: The background image is taken with {\it Spitzer} MIPS $24~\mu m$ image. The image pixel has the size of $2.49\arcsec\times 2.60\arcsec$. We use MIPS photometry for absolute flux calibration for the IRS spectrum.} \label{fig:diskPA}
\end{figure}
\end{appendix}

\clearpage
\bibliographystyle{aas63_bren}
\bibliography{main}

\begin{thebibliography}{}
\expandafter\ifx\csname natexlab\endcsname\relax\def\natexlab#1{#1}\fi
\providecommand{\url}[1]{\href{#1}{#1}}
\providecommand{\dodoi}[1]{}
\providecommand{\doarXiv}[1]{\href{https://arxiv.org/abs/#1}{\nolinkurl{https://arxiv.org/abs/#1}}}

\bibitem[{{Allard} {et~al.}(2012){Allard}, {Homeier}, \& {Freytag}}]{Allard+12}
{Allard}, F., {Homeier}, D., \& {Freytag}, B. 2012,
  \href{http://dx.doi.org/10.1098/rsta.2011.0269}{\color{magenta}RSPTA},
  \href{https://ui.adsabs.harvard.edu/abs/2012RSPTA.370.2765A}{\color{blue}370},
  \href{https://ui.adsabs.harvard.edu/abs/2012RSPTA.370.2765A}{\color{blue}2765}

\bibitem[{{Ballering} {et~al.}(2016){Ballering}, {Su}, {Rieke}, \&
  {G{\'a}sp{\'a}r}}]{Ballering+16}
{Ballering}, N.~P., {Su}, K. Y.~L., {Rieke}, G.~H., \& {G{\'a}sp{\'a}r}, A.
  2016,
  \href{http://dx.doi.org/10.3847/0004-637X/823/2/108}{\color{magenta}\apj},
  \href{https://ui.adsabs.harvard.edu/abs/2016ApJ...823..108B}{\color{blue}823},
  \href{https://ui.adsabs.harvard.edu/abs/2016ApJ...823..108B}{\color{blue}108}

\bibitem[{{Bonnefoy} {et~al.}(2013){Bonnefoy}, {Boccaletti}, {Lagrange},
  {Allard}, {Mordasini}, {Beust}, {Chauvin}, {Girard}, {Homeier}, {Apai},
  {Lacour}, \& {Rouan}}]{Bonnefoy+13}
{Bonnefoy}, M., {Boccaletti}, A., {Lagrange}, A.~M., {et~al.} 2013,
  \href{http://dx.doi.org/10.1051/0004-6361/201220838}{\color{magenta}\aap},
  \href{https://ui.adsabs.harvard.edu/abs/2013A&A...555A.107B}{\color{blue}555},
  \href{https://ui.adsabs.harvard.edu/abs/2013A&A...555A.107B}{\color{blue}A107}

\bibitem[{{Bouchet} {et~al.}(1991){Bouchet}, {Manfroid}, \&
  {Schmider}}]{Bouchet+91}
{Bouchet}, P., {Manfroid}, J., \& {Schmider}, F.~X. 1991, \aaps,
  \href{https://ui.adsabs.harvard.edu/abs/1991A&AS...91..409B}{\color{blue}91},
  \href{https://ui.adsabs.harvard.edu/abs/1991A&AS...91..409B}{\color{blue}409}

\bibitem[{{Brownlee}(2008)}]{Brownlee+08}
{Brownlee}, D. 2008,
  \href{http://dx.doi.org/10.1063/1.2947646}{\color{magenta}Physics Today},
  \href{https://ui.adsabs.harvard.edu/abs/2008PhT....61f..30B}{\color{blue}61},
  \href{https://ui.adsabs.harvard.edu/abs/2008PhT....61f..30B}{\color{blue}30}

\bibitem[{{Chen} {et~al.}(2014){Chen}, {Mittal}, {Kuchner}, {Forrest}, {Lisse},
  {Manoj}, {Sargent}, \& {Watson}}]{Chen+14}
{Chen}, C.~H., {Mittal}, T., {Kuchner}, M., {et~al.} 2014,
  \href{http://dx.doi.org/10.1088/0067-0049/211/2/25}{\color{magenta}\apjs},
  \href{https://ui.adsabs.harvard.edu/abs/2014ApJS..211...25C}{\color{blue}211},
  \href{https://ui.adsabs.harvard.edu/abs/2014ApJS..211...25C}{\color{blue}25}

\bibitem[{{Chen} {et~al.}(2020){Chen}, {Su}, \& {Xu}}]{Chen+20}
{Chen}, C.~H., {Su}, K. Y.~L., \& {Xu}, S. 2020,
  \href{http://dx.doi.org/10.1038/s41550-020-1067-6}{\color{magenta}NatAs},
  \href{https://ui.adsabs.harvard.edu/abs/2020NatAs...4..328C}{\color{blue}4},
  \href{https://ui.adsabs.harvard.edu/abs/2020NatAs...4..328C}{\color{blue}328}

\bibitem[{{Chen} {et~al.}(2007){Chen}, {Li}, {Bohac}, {Kim}, {Watson}, {van
  Cleve}, {Houck}, {Stapelfeldt}, {Werner}, {Rieke}, {Su}, {Marengo},
  {Backman}, {Beichman}, \& {Fazio}}]{Chen+07}
{Chen}, C.~H., {Li}, A., {Bohac}, C., {et~al.} 2007,
  \href{http://dx.doi.org/10.1086/519989}{\color{magenta}\apj},
  \href{https://ui.adsabs.harvard.edu/abs/2007ApJ...666..466C}{\color{blue}666},
  \href{https://ui.adsabs.harvard.edu/abs/2007ApJ...666..466C}{\color{blue}466}

\bibitem[{{Chihara} {et~al.}(2002){Chihara}, {Koike}, {Tsuchiyama},
  {Tachibana}, \& {Sakamoto}}]{Chihara+02}
{Chihara}, H., {Koike}, C., {Tsuchiyama}, A., {et~al.} 2002,
  \href{http://dx.doi.org/10.1051/0004-6361:20020791}{\color{magenta}\aap},
  \href{https://ui.adsabs.harvard.edu/abs/2002A&A...391..267C}{\color{blue}391},
  \href{https://ui.adsabs.harvard.edu/abs/2002A&A...391..267C}{\color{blue}267}

\bibitem[{{Claret}(2000)}]{Claret2000}
{Claret}, A. 2000, \aap,
  \href{https://ui.adsabs.harvard.edu/abs/2000A&A...363.1081C}{\color{blue}363},
  \href{https://ui.adsabs.harvard.edu/abs/2000A&A...363.1081C}{\color{blue}1081}

\bibitem[{{Claret} {et~al.}(1995){Claret}, {Diaz-Cordoves}, \&
  {Gimenez}}]{Claret+95}
{Claret}, A., {Diaz-Cordoves}, J., \& {Gimenez}, A. 1995, \aaps,
  \href{https://ui.adsabs.harvard.edu/abs/1995A&AS..114..247C}{\color{blue}114},
  \href{https://ui.adsabs.harvard.edu/abs/1995A&AS..114..247C}{\color{blue}247}

\bibitem[{{Cushing} {et~al.}(2004){Cushing}, {Vacca}, \& {Rayner}}]{Cushing+04}
{Cushing}, M.~C., {Vacca}, W.~D., \& {Rayner}, J.~T. 2004,
  \href{http://dx.doi.org/10.1086/382907}{\color{magenta}\pasp},
  \href{https://ui.adsabs.harvard.edu/abs/2004PASP..116..362C}{\color{blue}116},
  \href{https://ui.adsabs.harvard.edu/abs/2004PASP..116..362C}{\color{blue}362}

\bibitem[{{Cutri} \& {et al.}(2012)}]{Cutri+12}
{Cutri}, R.~M., \& {et al.} 2012, VizieR Online Data Catalog,
  \href{https://ui.adsabs.harvard.edu/abs/2012yCat.2311....0C}{\color{blue}II/311}

\bibitem[{{Czechowski} \& {Mann}(2007)}]{Czechowski+Mann07}
{Czechowski}, A., \& {Mann}, I. 2007,
  \href{http://dx.doi.org/10.1086/512965}{\color{magenta}\apj},
  \href{https://ui.adsabs.harvard.edu/abs/2007ApJ...660.1541C}{\color{blue}660},
  \href{https://ui.adsabs.harvard.edu/abs/2007ApJ...660.1541C}{\color{blue}1541}

\bibitem[{{de Vries} {et~al.}(2012){de Vries}, {Acke}, {Blommaert}, {Waelkens},
  {Waters}, {Vandenbussche}, {Min}, {Olofsson}, {Dominik}, {Decin}, {Barlow},
  {Brandeker}, {di Francesco}, {Glauser}, {Greaves}, {Harvey}, {Holland},
  {Ivison}, {Liseau}, {Pantin}, {Pilbratt}, {Royer}, \&
  {Sibthorpe}}]{deVries+12}
{de Vries}, B.~L., {Acke}, B., {Blommaert}, J.~A.~D.~L., {et~al.} 2012,
  \href{http://dx.doi.org/10.1038/nature11469}{\color{magenta}\nat},
  \href{https://ui.adsabs.harvard.edu/abs/2012Natur.490...74D}{\color{blue}490},
  \href{https://ui.adsabs.harvard.edu/abs/2012Natur.490...74D}{\color{blue}74}

\bibitem[{{Defr{\`e}re} {et~al.}(2012){Defr{\`e}re}, {Lebreton}, {Le Bouquin},
  {Lagrange}, {Absil}, {Augereau}, {Berger}, {di Folco}, {Ertel}, {Kluska},
  {Montagnier}, {Millan-Gabet}, {Traub}, \& {Zins}}]{Defrere+12}
{Defr{\`e}re}, D., {Lebreton}, J., {Le Bouquin}, J.~B., {et~al.} 2012,
  \href{http://dx.doi.org/10.1051/0004-6361/201220287}{\color{magenta}\aap},
  \href{https://ui.adsabs.harvard.edu/abs/2012A&A...546L...9D}{\color{blue}546},
  \href{https://ui.adsabs.harvard.edu/abs/2012A&A...546L...9D}{\color{blue}L9}

\bibitem[{DeMeo {et~al.}(2019)DeMeo, Polishook, Carry, Burt, Hsieh, Binzel,
  Moskovitz, \& Burbine}]{demeo+2019}
DeMeo, F.~E., Polishook, D., Carry, B., {et~al.} 2019, Icarus, 322, 13

\bibitem[{{Dent} {et~al.}(2014){Dent}, {Wyatt}, {Roberge}, {Augereau},
  {Casassus}, {Corder}, {Greaves}, {de Gregorio-Monsalvo}, {Hales}, {Jackson},
  {Hughes}, {Lagrange}, {Matthews}, \& {Wilner}}]{Dent+14}
{Dent}, W.~R.~F., {Wyatt}, M.~C., {Roberge}, A., {et~al.} 2014,
  \href{http://dx.doi.org/10.1126/science.1248726}{\color{magenta}Science},
  \href{https://ui.adsabs.harvard.edu/abs/2014Sci...343.1490D}{\color{blue}343},
  \href{https://ui.adsabs.harvard.edu/abs/2014Sci...343.1490D}{\color{blue}1490}

\bibitem[{{Dohnanyi}(1969)}]{Dohnanyi+69}
{Dohnanyi}, J.~S. 1969,
  \href{http://dx.doi.org/10.1029/JB074i010p02531}{\color{magenta}JGR},
  \href{https://ui.adsabs.harvard.edu/abs/1969JGR....74.2531D}{\color{blue}74},
  \href{https://ui.adsabs.harvard.edu/abs/1969JGR....74.2531D}{\color{blue}2531}

\bibitem[{{Dorschner} {et~al.}(1995){Dorschner}, {Begemann}, {Henning},
  {Jaeger}, \& {Mutschke}}]{Dorschner+95}
{Dorschner}, J., {Begemann}, B., {Henning}, T., {et~al.} 1995, \aap,
  \href{https://ui.adsabs.harvard.edu/abs/1995A&A...300..503D}{\color{blue}300},
  \href{https://ui.adsabs.harvard.edu/abs/1995A&A...300..503D}{\color{blue}503}

\bibitem[{{Ducati}(2002)}]{Ducati02}
{Ducati}, J.~R. 2002, VizieR Online Data Catalog,
  \href{https://ui.adsabs.harvard.edu/abs/2002yCat.2237....0D}{\color{blue}II/237}

\bibitem[{{Ertel} {et~al.}(2014){Ertel}, {Absil}, {Defr{\`e}re}, {Le Bouquin},
  {Augereau}, {Marion}, {Blind}, {Bonsor}, {Bryden}, {Lebreton}, \&
  {Milli}}]{Ertel+14}
{Ertel}, S., {Absil}, O., {Defr{\`e}re}, D., {et~al.} 2014,
  \href{http://dx.doi.org/10.1051/0004-6361/201424438}{\color{magenta}\aap},
  \href{https://ui.adsabs.harvard.edu/abs/2014A&A...570A.128E}{\color{blue}570},
  \href{https://ui.adsabs.harvard.edu/abs/2014A&A...570A.128E}{\color{blue}A128}

\bibitem[{{Fabian} {et~al.}(2001){Fabian}, {Henning}, {J{\"a}ger}, {Mutschke},
  {Dorschner}, \& {Wehrhan}}]{Fabian+01}
{Fabian}, D., {Henning}, T., {J{\"a}ger}, C., {et~al.} 2001,
  \href{http://dx.doi.org/10.1051/0004-6361:20011196}{\color{magenta}\aap},
  \href{https://ui.adsabs.harvard.edu/abs/2001A&A...378..228F}{\color{blue}378},
  \href{https://ui.adsabs.harvard.edu/abs/2001A&A...378..228F}{\color{blue}228}

\bibitem[{{Fujiwara} {et~al.}(2010){Fujiwara}, {Onaka}, {Ishihara},
  {Yamashita}, {Fukagawa}, {Nakagawa}, {Kataza}, {Ootsubo}, \&
  {Murakami}}]{Fujiwara+10}
{Fujiwara}, H., {Onaka}, T., {Ishihara}, D., {et~al.} 2010,
  \href{http://dx.doi.org/10.1088/2041-8205/714/1/L152}{\color{magenta}\apjl},
  \href{https://ui.adsabs.harvard.edu/abs/2010ApJ...714L.152F}{\color{blue}714},
  \href{https://ui.adsabs.harvard.edu/abs/2010ApJ...714L.152F}{\color{blue}L152}

\bibitem[{{Guess}(1962)}]{Guess62}
{Guess}, A.~W. 1962,
  \href{http://dx.doi.org/10.1086/147329}{\color{magenta}\apj},
  \href{https://ui.adsabs.harvard.edu/abs/1962ApJ...135..855G}{\color{blue}135},
  \href{https://ui.adsabs.harvard.edu/abs/1962ApJ...135..855G}{\color{blue}855}

\bibitem[{Hamilton(2010)}]{Hamilton10}
Hamilton, V.~E. 2010, Geochemistry, 70, 7

\bibitem[{{Hauschildt} {et~al.}(1999){Hauschildt}, {Allard}, \&
  {Baron}}]{Hauschildt+99}
{Hauschildt}, P.~H., {Allard}, F., \& {Baron}, E. 1999,
  \href{http://dx.doi.org/10.1086/306745}{\color{magenta}\apj},
  \href{https://ui.adsabs.harvard.edu/abs/1999ApJ...512..377H}{\color{blue}512},
  \href{https://ui.adsabs.harvard.edu/abs/1999ApJ...512..377H}{\color{blue}377}

\bibitem[{{Henning}(2010)}]{Henning2010}
{Henning}, T. 2010,
  \href{http://dx.doi.org/10.1146/annurev-astro-081309-130815}{\color{magenta}\araa},
  \href{https://ui.adsabs.harvard.edu/abs/2010ARA&A..48...21H}{\color{blue}48},
  \href{https://ui.adsabs.harvard.edu/abs/2010ARA&A..48...21H}{\color{blue}21}

\bibitem[{{Higdon} {et~al.}(2004){Higdon}, {Devost}, {Higdon}, {Brand l},
  {Houck}, {Hall}, {Barry}, {Charmand aris}, {Smith}, {Sloan}, \&
  {Green}}]{Higdon+04}
{Higdon}, S.~J.~U., {Devost}, D., {Higdon}, J.~L., {et~al.} 2004,
  \href{http://dx.doi.org/10.1086/425083}{\color{magenta}\pasp},
  \href{https://ui.adsabs.harvard.edu/abs/2004PASP..116..975H}{\color{blue}116},
  \href{https://ui.adsabs.harvard.edu/abs/2004PASP..116..975H}{\color{blue}975}

\bibitem[{{Houck} {et~al.}(2004){Houck}, {Roellig}, {van Cleve}, {Forrest},
  {Herter}, {Lawrence}, {Matthews}, {Reitsema}, {Soifer}, {Watson}, {Weedman},
  {Huisjen}, {Troeltzsch}, {Barry}, {Bernard-Salas}, {Blacken}, {Brandl},
  {Charmandaris}, {Devost}, {Gull}, {Hall}, {Henderson}, {Higdon}, {Pirger},
  {Schoenwald}, {Sloan}, {Uchida}, {Appleton}, {Armus}, {Burgdorf},
  {Fajardo-Acosta}, {Grillmair}, {Ingalls}, {Morris}, \& {Teplitz}}]{Houck+04}
{Houck}, J.~R., {Roellig}, T.~L., {van Cleve}, J., {et~al.} 2004,
  \href{http://dx.doi.org/10.1086/423134}{\color{magenta}\apjs},
  \href{https://ui.adsabs.harvard.edu/abs/2004ApJS..154...18H}{\color{blue}154},
  \href{https://ui.adsabs.harvard.edu/abs/2004ApJS..154...18H}{\color{blue}18}

\bibitem[{{Hughes} {et~al.}(2018){Hughes}, {Duch{\^e}ne}, \&
  {Matthews}}]{Hughes+18}
{Hughes}, A.~M., {Duch{\^e}ne}, G., \& {Matthews}, B.~C. 2018,
  \href{http://dx.doi.org/10.1146/annurev-astro-081817-052035}{\color{magenta}\araa},
  \href{https://ui.adsabs.harvard.edu/abs/2018ARA&A..56..541H}{\color{blue}56},
  \href{https://ui.adsabs.harvard.edu/abs/2018ARA&A..56..541H}{\color{blue}541}

\bibitem[{{Kessler-Silacci} {et~al.}(2006){Kessler-Silacci}, {Augereau},
  {Dullemond}, {Geers}, {Lahuis}, {Evans}, {van Dishoeck}, {Blake}, {Boogert},
  {Brown}, {J{\o}rgensen}, {Knez}, \& {Pontoppidan}}]{Kessler-Silacci+06}
{Kessler-Silacci}, J., {Augereau}, J.-C., {Dullemond}, C.~P., {et~al.} 2006,
  \href{http://dx.doi.org/10.1086/499330}{\color{magenta}\apj},
  \href{https://ui.adsabs.harvard.edu/abs/2006ApJ...639..275K}{\color{blue}639},
  \href{https://ui.adsabs.harvard.edu/abs/2006ApJ...639..275K}{\color{blue}275}

\bibitem[{{Kiefer} {et~al.}(2014){Kiefer}, {Lecavelier des Etangs}, {Boissier},
  {Vidal-Madjar}, {Beust}, {Lagrange}, {H{\'e}brard}, \& {Ferlet}}]{Kiefer+14}
{Kiefer}, F., {Lecavelier des Etangs}, A., {Boissier}, J., {et~al.} 2014,
  \href{http://dx.doi.org/10.1038/nature13849}{\color{magenta}\nat},
  \href{https://ui.adsabs.harvard.edu/abs/2014Natur.514..462K}{\color{blue}514},
  \href{https://ui.adsabs.harvard.edu/abs/2014Natur.514..462K}{\color{blue}462}

\bibitem[{{Koike} {et~al.}(2003){Koike}, {Chihara}, {Tsuchiyama}, {Suto},
  {Sogawa}, \& {Okuda}}]{Koike+03}
{Koike}, C., {Chihara}, H., {Tsuchiyama}, A., {et~al.} 2003,
  \href{http://dx.doi.org/10.1051/0004-6361:20021831}{\color{magenta}\aap},
  \href{https://ui.adsabs.harvard.edu/abs/2003A&A...399.1101K}{\color{blue}399},
  \href{https://ui.adsabs.harvard.edu/abs/2003A&A...399.1101K}{\color{blue}1101}

\bibitem[{{Kolokolova} \& {Kimura}(2010)}]{Kolokolova+10}
{Kolokolova}, L., \& {Kimura}, H. 2010,
  \href{http://dx.doi.org/10.5047/eps.2008.12.001}{\color{magenta}Earth,
  Planets and Space},
  \href{https://ui.adsabs.harvard.edu/abs/2010EP&S...62...17K}{\color{blue}62},
  \href{https://ui.adsabs.harvard.edu/abs/2010EP&S...62...17K}{\color{blue}17}

\bibitem[{{Kral} {et~al.}(2016){Kral}, {Wyatt}, {Carswell}, {Pringle},
  {Matr{\`a}}, \& {Juh{\'a}sz}}]{Kral+16}
{Kral}, Q., {Wyatt}, M., {Carswell}, R.~F., {et~al.} 2016,
  \href{http://dx.doi.org/10.1093/mnras/stw1361}{\color{magenta}\mnras},
  \href{https://ui.adsabs.harvard.edu/abs/2016MNRAS.461..845K}{\color{blue}461},
  \href{https://ui.adsabs.harvard.edu/abs/2016MNRAS.461..845K}{\color{blue}845}

\bibitem[{{Lagrange} {et~al.}(2009){Lagrange}, {Gratadour}, {Chauvin}, {Fusco},
  {Ehrenreich}, {Mouillet}, {Rousset}, {Rouan}, {Allard}, {Gendron}, {Charton},
  {Mugnier}, {Rabou}, {Montri}, \& {Lacombe}}]{Lagrange+09}
{Lagrange}, A.~M., {Gratadour}, D., {Chauvin}, G., {et~al.} 2009,
  \href{http://dx.doi.org/10.1051/0004-6361:200811325}{\color{magenta}\aap},
  \href{https://ui.adsabs.harvard.edu/abs/2009A&A...493L..21L}{\color{blue}493},
  \href{https://ui.adsabs.harvard.edu/abs/2009A&A...493L..21L}{\color{blue}L21}

\bibitem[{{Lagrange} {et~al.}(2010){Lagrange}, {Bonnefoy}, {Chauvin}, {Apai},
  {Ehrenreich}, {Boccaletti}, {Gratadour}, {Rouan}, {Mouillet}, {Lacour}, \&
  {Kasper}}]{Lagrange+10}
{Lagrange}, A.~M., {Bonnefoy}, M., {Chauvin}, G., {et~al.} 2010,
  \href{http://dx.doi.org/10.1126/science.1187187}{\color{magenta}Science},
  \href{https://ui.adsabs.harvard.edu/abs/2010Sci...329...57L}{\color{blue}329},
  \href{https://ui.adsabs.harvard.edu/abs/2010Sci...329...57L}{\color{blue}57}

\bibitem[{{Lagrange} {et~al.}(2020){Lagrange}, {Rubini}, {Nowak}, {Lacour},
  {Grandjean}, {Boccaletti}, {Langlois}, {Delorme}, {Gratton}, {Wang},
  {Flasseur}, {Galicher}, {Kral}, {Meunier}, {Beust}, {Babusiaux}, {Le
  Coroller}, {Thebault}, {Kervella}, {Zurlo}, {Maire}, {Wahhaj}, {Amorim},
  {Asensio-Torres}, {Benisty}, {Berger}, {Bonnefoy}, {Brandner}, {Cantalloube},
  {Charnay}, {Chauvin}, {Choquet}, {Cl{\'e}net}, {Christiaens}, {Coud{\'e} Du
  Foresto}, {de Zeeuw}, {Desidera}, {Duvert}, {Eckart}, {Eisenhauer},
  {Galland}, {Gao}, {Garcia}, {Garcia Lopez}, {Gendron}, {Genzel}, {Gillessen},
  {Girard}, {Hagelberg}, {Haubois}, {Henning}, {Heissel}, {Hippler},
  {Horrobin}, {Janson}, {Kammerer}, {Kenworthy}, {Keppler}, {Kreidberg},
  {Lapeyr{\`e}re}, {Le Bouquin}, {L{\'e}na}, {M{\'e}rand}, {Messina},
  {Molli{\`e}re}, {Monnier}, {Ott}, {Otten}, {Paumard}, {Paladini}, {Perraut},
  {Perrin}, {Pueyo}, {Pfuhl}, {Rodet}, {Rodriguez-Coira}, {Rousset}, {Samland},
  {Shangguan}, {Schmidt}, {Straub}, {Straubmeier}, {Stolker}, {Vigan},
  {Vincent}, {Widmann}, {Woillez}, \& {Gravity Collaboration}}]{Lagrange+20}
{Lagrange}, A.~M., {Rubini}, P., {Nowak}, M., {et~al.} 2020,
  \href{http://dx.doi.org/10.1051/0004-6361/202038823}{\color{magenta}\aap},
  \href{https://ui.adsabs.harvard.edu/abs/2020A&A...642A..18L}{\color{blue}642},
  \href{https://ui.adsabs.harvard.edu/abs/2020A&A...642A..18L}{\color{blue}A18}

\bibitem[{{Larwood} \& {Kalas}(2001)}]{Larwood+01}
{Larwood}, J.~D., \& {Kalas}, P.~G. 2001,
  \href{http://dx.doi.org/10.1046/j.1365-8711.2001.04212.x}{\color{magenta}\mnras},
  \href{https://ui.adsabs.harvard.edu/abs/2001MNRAS.323..402L}{\color{blue}323},
  \href{https://ui.adsabs.harvard.edu/abs/2001MNRAS.323..402L}{\color{blue}402}

\bibitem[{{Le Guillou} {et~al.}(2015){Le Guillou}, Changela, \&
  Brearley}]{LeGuillou+15}
{Le Guillou}, C., Changela, H.~G., \& Brearley, A.~J. 2015,
  \href{http://dx.doi.org/https://doi.org/10.1016/j.epsl.2015.02.031}{\color{magenta}Earth
  and Planetary Science Letters}, 420, 162

\bibitem[{{Lebouteiller} {et~al.}(2010){Lebouteiller}, {Bernard-Salas},
  {Sloan}, \& {Barry}}]{Lebouteiller+10}
{Lebouteiller}, V., {Bernard-Salas}, J., {Sloan}, G.~C., \& {Barry}, D.~J.
  2010, \href{http://dx.doi.org/10.1086/650426}{\color{magenta}\pasp},
  \href{https://ui.adsabs.harvard.edu/abs/2010PASP..122..231L}{\color{blue}122},
  \href{https://ui.adsabs.harvard.edu/abs/2010PASP..122..231L}{\color{blue}231}

\bibitem[{{Li} \& {Greenberg}(1998)}]{Li+Greenberg98}
{Li}, A., \& {Greenberg}, J.~M. 1998, \aap,
  \href{https://ui.adsabs.harvard.edu/abs/1998A&A...331..291L}{\color{blue}331},
  \href{https://ui.adsabs.harvard.edu/abs/1998A&A...331..291L}{\color{blue}291}

\bibitem[{{Lisse} {et~al.}(2007){Lisse}, {Kraemer}, {Nuth}, {Li}, \&
  {Joswiak}}]{Lisse+07Comp}
{Lisse}, C.~M., {Kraemer}, K.~E., {Nuth}, J.~A., {et~al.} 2007,
  \href{http://dx.doi.org/10.1016/j.icarus.2006.11.019}{\color{magenta}\icarus},
  \href{https://ui.adsabs.harvard.edu/abs/2007Icar..187...69L}{\color{blue}187},
  \href{https://ui.adsabs.harvard.edu/abs/2007Icar..187...69L}{\color{blue}69}

\bibitem[{{Lisse} {et~al.}(2006){Lisse}, {VanCleve}, {Adams}, {A'Hearn},
  {Fern{\'a}ndez}, {Farnham}, {Armus}, {Grillmair}, {Ingalls}, {Belton},
  {Groussin}, {McFadden}, {Meech}, {Schultz}, {Clark}, {Feaga}, \&
  {Sunshine}}]{Lisse+06}
{Lisse}, C.~M., {VanCleve}, J., {Adams}, A.~C., {et~al.} 2006,
  \href{http://dx.doi.org/10.1126/science.1124694}{\color{magenta}Science},
  \href{https://ui.adsabs.harvard.edu/abs/2006Sci...313..635L}{\color{blue}313},
  \href{https://ui.adsabs.harvard.edu/abs/2006Sci...313..635L}{\color{blue}635}

\bibitem[{{Lytle} {et~al.}(1999){Lytle}, {Stobie}, {Ferro}, \&
  {Barg}}]{Lytle+99}
{Lytle}, D., {Stobie}, E., {Ferro}, A., \& {Barg}, I. 1999, Astronomical
  Society of the Pacific Conference Series,
  \href{https://ui.adsabs.harvard.edu/abs/1999ASPC..172..445L}{\color{blue}172},
  \href{https://ui.adsabs.harvard.edu/abs/1999ASPC..172..445L}{\color{blue}445}

\bibitem[{{Mainzer} {et~al.}(2011){Mainzer}, {Bauer}, {Grav}, {Masiero},
  {Cutri}, {Dailey}, {Eisenhardt}, {McMillan}, {Wright}, {Walker}, {Jedicke},
  {Spahr}, {Tholen}, {Alles}, {Beck}, {Brandenburg}, {Conrow}, {Evans},
  {Fowler}, {Jarrett}, {Marsh}, {Masci}, {McCallon}, {Wheelock}, {Wittman},
  {Wyatt}, {DeBaun}, {Elliott}, {Elsbury}, {Gautier}, {Gomillion}, {Leisawitz},
  {Maleszewski}, {Micheli}, \& {Wilkins}}]{Mainzer+11}
{Mainzer}, A., {Bauer}, J., {Grav}, T., {et~al.} 2011,
  \href{http://dx.doi.org/10.1088/0004-637X/731/1/53}{\color{magenta}\apj},
  \href{https://ui.adsabs.harvard.edu/abs/2011ApJ...731...53M}{\color{blue}731},
  \href{https://ui.adsabs.harvard.edu/abs/2011ApJ...731...53M}{\color{blue}53}

\bibitem[{{Mermilliod} {et~al.}(1997){Mermilliod}, {Mermilliod}, \&
  {Hauck}}]{Mermilliod+97}
{Mermilliod}, J.~C., {Mermilliod}, M., \& {Hauck}, B. 1997,
  \href{http://dx.doi.org/10.1051/aas:1997197}{\color{magenta}\aaps},
  \href{https://ui.adsabs.harvard.edu/abs/1997A&AS..124..349M}{\color{blue}124},
  \href{https://ui.adsabs.harvard.edu/abs/1997A&AS..124..349M}{\color{blue}349}

\bibitem[{{Mittal} {et~al.}(2015){Mittal}, {Chen}, {Jang-Condell}, {Manoj},
  {Sargent}, {Watson}, \& {Lisse}}]{Mittal+15}
{Mittal}, T., {Chen}, C.~H., {Jang-Condell}, H., {et~al.} 2015,
  \href{http://dx.doi.org/10.1088/0004-637X/798/2/87}{\color{magenta}\apj},
  \href{https://ui.adsabs.harvard.edu/abs/2015ApJ...798...87M}{\color{blue}798},
  \href{https://ui.adsabs.harvard.edu/abs/2015ApJ...798...87M}{\color{blue}87}

\bibitem[{{Morales} {et~al.}(2012){Morales}, {Padgett}, {Bryden}, {Werner}, \&
  {Furlan}}]{Morales+12}
{Morales}, F.~Y., {Padgett}, D.~L., {Bryden}, G., {et~al.} 2012,
  \href{http://dx.doi.org/10.1088/0004-637X/757/1/7}{\color{magenta}\apj},
  \href{https://ui.adsabs.harvard.edu/abs/2012ApJ...757....7M}{\color{blue}757},
  \href{https://ui.adsabs.harvard.edu/abs/2012ApJ...757....7M}{\color{blue}7}

\bibitem[{{Nowak} {et~al.}(2020){Nowak}, {Lacour}, {Lagrange}, {Rubini},
  {Wang}, {Stolker}, {Abuter}, {Amorim}, {Asensio-Torres}, {Baub{\"o}ck},
  {Benisty}, {Berger}, {Beust}, {Blunt}, {Boccaletti}, {Bonnefoy}, {Bonnet},
  {Brandner}, {Cantalloube}, {Charnay}, {Choquet}, {Christiaens}, {Cl{\'e}net},
  {Coud{\'e} Du Foresto}, {Cridland}, {de Zeeuw}, {Dembet}, {Dexter},
  {Drescher}, {Duvert}, {Eckart}, {Eisenhauer}, {Gao}, {Garcia}, {Garcia
  Lopez}, {Gardner}, {Gendron}, {Genzel}, {Gillessen}, {Girard}, {Grandjean},
  {Haubois}, {Hei{\ss}el}, {Henning}, {Hinkley}, {Hippler}, {Horrobin},
  {Houll{\'e}}, {Hubert}, {Jim{\'e}nez-Rosales}, {Jocou}, {Kammerer},
  {Kervella}, {Keppler}, {Kreidberg}, {Kulikauskas}, {Lapeyr{\`e}re}, {Le
  Bouquin}, {L{\'e}na}, {M{\'e}rand}, {Maire}, {Molli{\`e}re}, {Monnier},
  {Mouillet}, {M{\"u}ller}, {Nasedkin}, {Ott}, {Otten}, {Paumard}, {Paladini},
  {Perraut}, {Perrin}, {Pueyo}, {Pfuhl}, {Rameau}, {Rodet},
  {Rodr{\'\i}guez-Coira}, {Rousset}, {Scheithauer}, {Shangguan}, {Stadler},
  {Straub}, {Straubmeier}, {Sturm}, {Tacconi}, {van Dishoeck}, {Vigan},
  {Vincent}, {von Fellenberg}, {Ward-Duong}, {Widmann}, {Wieprecht},
  {Wiezorrek}, {Woillez}, \& {Gravity Collaboration}}]{Nowak+20}
{Nowak}, M., {Lacour}, S., {Lagrange}, A.~M., {et~al.} 2020,
  \href{http://dx.doi.org/10.1051/0004-6361/202039039}{\color{magenta}\aap},
  \href{https://ui.adsabs.harvard.edu/abs/2020A&A...642L...2N}{\color{blue}642},
  \href{https://ui.adsabs.harvard.edu/abs/2020A&A...642L...2N}{\color{blue}L2}

\bibitem[{{Ochsenbein} {et~al.}(2000){Ochsenbein}, {Bauer}, \&
  {Marcout}}]{Ochsenbein+2000}
{Ochsenbein}, F., {Bauer}, P., \& {Marcout}, J. 2000,
  \href{http://dx.doi.org/10.1051/aas:2000169}{\color{magenta}\aaps},
  \href{https://ui.adsabs.harvard.edu/abs/2000A&AS..143...23O}{\color{blue}143},
  \href{https://ui.adsabs.harvard.edu/abs/2000A&AS..143...23O}{\color{blue}23}

\bibitem[{{Okamoto} {et~al.}(2004){Okamoto}, {Kataza}, {Honda}, {Yamashita},
  {Onaka}, {Watanabe}, {Miyata}, {Sako}, {Fujiyoshi}, \& {Sakon}}]{Okamoto+04}
{Okamoto}, Y.~K., {Kataza}, H., {Honda}, M., {et~al.} 2004,
  \href{http://dx.doi.org/10.1038/nature02948}{\color{magenta}\nat},
  \href{https://ui.adsabs.harvard.edu/abs/2004Natur.431..660O}{\color{blue}431},
  \href{https://ui.adsabs.harvard.edu/abs/2004Natur.431..660O}{\color{blue}660}

\bibitem[{{Olofsson} {et~al.}(2009){Olofsson}, {Augereau}, {van Dishoeck},
  {Mer{\'\i}n}, {Lahuis}, {Kessler-Silacci}, {Dullemond}, {Oliveira}, {Blake},
  {Boogert}, {Brown}, {Evans}, {Geers}, {Knez}, {Monin}, \&
  {Pontoppidan}}]{Olofsson+09}
{Olofsson}, J., {Augereau}, J.~C., {van Dishoeck}, E.~F., {et~al.} 2009,
  \href{http://dx.doi.org/10.1051/0004-6361/200912062}{\color{magenta}\aap},
  \href{https://ui.adsabs.harvard.edu/abs/2009A&A...507..327O}{\color{blue}507},
  \href{https://ui.adsabs.harvard.edu/abs/2009A&A...507..327O}{\color{blue}327}

\bibitem[{{Pan} \& {Schlichting}(2012)}]{Pan+Schlichting12}
{Pan}, M., \& {Schlichting}, H.~E. 2012,
  \href{http://dx.doi.org/10.1088/0004-637X/747/2/113}{\color{magenta}\apj},
  \href{https://ui.adsabs.harvard.edu/abs/2012ApJ...747..113P}{\color{blue}747},
  \href{https://ui.adsabs.harvard.edu/abs/2012ApJ...747..113P}{\color{blue}113}

\bibitem[{{Pecaut} \& {Mamajek}(2013)}]{Pecaut+Mamajek13}
{Pecaut}, M.~J., \& {Mamajek}, E.~E. 2013,
  \href{http://dx.doi.org/10.1088/0067-0049/208/1/9}{\color{magenta}\apjs},
  \href{https://ui.adsabs.harvard.edu/abs/2013ApJS..208....9P}{\color{blue}208},
  \href{https://ui.adsabs.harvard.edu/abs/2013ApJS..208....9P}{\color{blue}9}

\bibitem[{{Rayner} {et~al.}(2003){Rayner}, {Toomey}, {Onaka}, {Denault},
  {Stahlberger}, {Vacca}, {Cushing}, \& {Wang}}]{Rayner+2003}
{Rayner}, J.~T., {Toomey}, D.~W., {Onaka}, P.~M., {et~al.} 2003,
  \href{http://dx.doi.org/10.1086/367745}{\color{magenta}\pasp},
  \href{https://ui.adsabs.harvard.edu/abs/2003PASP..115..362R}{\color{blue}115},
  \href{https://ui.adsabs.harvard.edu/abs/2003PASP..115..362R}{\color{blue}362}

\bibitem[{{Reach} {et~al.}(2010){Reach}, {Vaubaillon}, {Lisse}, {Holloway}, \&
  {Rho}}]{Reach+10}
{Reach}, W.~T., {Vaubaillon}, J., {Lisse}, C.~M., {et~al.} 2010,
  \href{http://dx.doi.org/10.1016/j.icarus.2010.01.020}{\color{magenta}\icarus},
  \href{https://ui.adsabs.harvard.edu/abs/2010Icar..208..276R}{\color{blue}208},
  \href{https://ui.adsabs.harvard.edu/abs/2010Icar..208..276R}{\color{blue}276}

\bibitem[{{Ren} {et~al.}(2017){Ren}, {Pueyo}, {Perrin}, {Debes}, \&
  {Choquet}}]{Ren+17}
{Ren}, B., {Pueyo}, L., {Perrin}, M.~D., {et~al.} 2017,
  \href{http://dx.doi.org/10.1117/12.2274163}{\color{magenta}Proc.~SPIE},
  \href{https://ui.adsabs.harvard.edu/abs/2017SPIE10400E..21R}{\color{blue}10400},
  \href{https://ui.adsabs.harvard.edu/abs/2017SPIE10400E..21R}{\color{blue}1040021}

\bibitem[{{Sargent} {et~al.}(2009{\natexlab{a}}){Sargent}, {Forrest},
  {Tayrien}, {McClure}, {Watson}, {Sloan}, {Li}, {Manoj}, {Bohac}, {Furlan},
  {Kim}, \& {Green}}]{Sargent+09b}
{Sargent}, B.~A., {Forrest}, W.~J., {Tayrien}, C., {et~al.} 2009{\natexlab{a}},
  \href{http://dx.doi.org/10.1088/0067-0049/182/2/477}{\color{magenta}\apjs},
  \href{https://ui.adsabs.harvard.edu/abs/2009ApJS..182..477S}{\color{blue}182},
  \href{https://ui.adsabs.harvard.edu/abs/2009ApJS..182..477S}{\color{blue}477}

\bibitem[{{Sargent} {et~al.}(2009{\natexlab{b}}){Sargent}, {Forrest},
  {Tayrien}, {McClure}, {Li}, {Basu}, {Manoj}, {Watson}, {Bohac}, {Furlan},
  {Kim}, {Green}, \& {Sloan}}]{Sargent+09ApJ}
---. 2009{\natexlab{b}},
  \href{http://dx.doi.org/10.1088/0004-637X/690/2/1193}{\color{magenta}\apj},
  \href{https://ui.adsabs.harvard.edu/abs/2009ApJ...690.1193S}{\color{blue}690},
  \href{https://ui.adsabs.harvard.edu/abs/2009ApJ...690.1193S}{\color{blue}1193}

\bibitem[{{Schneider} {et~al.}(2014){Schneider}, {Grady}, {Hines}, {Stark},
  {Debes}, {Carson}, {Kuchner}, {Perrin}, {Weinberger}, {Wisniewski},
  {Silverstone}, {Jang-Condell}, {Henning}, {Woodgate}, {Serabyn},
  {Moro-Martin}, {Tamura}, {Hinz}, \& {Rodigas}}]{Schneider+14}
{Schneider}, G., {Grady}, C.~A., {Hines}, D.~C., {et~al.} 2014,
  \href{http://dx.doi.org/10.1088/0004-6256/148/4/59}{\color{magenta}\aj},
  \href{https://ui.adsabs.harvard.edu/abs/2014AJ....148...59S}{\color{blue}148},
  \href{https://ui.adsabs.harvard.edu/abs/2014AJ....148...59S}{\color{blue}59}

\bibitem[{{Schneider} {et~al.}(2018){Schneider}, {Debes}, {Grady},
  {G{\'a}sp{\'a}r}, {Henning}, {Hines}, {Kuchner}, {Perrin}, \&
  {Wisniewski}}]{Schneider+18}
{Schneider}, G., {Debes}, J.~H., {Grady}, C.~A., {et~al.} 2018,
  \href{http://dx.doi.org/10.3847/1538-3881/aaa3f3}{\color{magenta}\aj},
  \href{https://ui.adsabs.harvard.edu/abs/2018AJ....155...77S}{\color{blue}155},
  \href{https://ui.adsabs.harvard.edu/abs/2018AJ....155...77S}{\color{blue}77}

\bibitem[{{Sings IRS Team} {et~al.}(2011){Sings IRS Team}, {Smith}, {Armus},
  {Bot}, {Buckalew}, {Dale}, {Helou}, {Jarrett}, {Roussel}, \&
  {Sheth}}]{IRScubism}
{Sings IRS Team}, {Smith}, J.~D., {Armus}, L., {et~al.} 2011, {CUBISM: CUbe
  Builder for IRS Spectra Maps}

\bibitem[{{Sloan} {et~al.}(2015){Sloan}, {Herter}, {Charmandaris}, {Sheth},
  {Burgdorf}, \& {Houck}}]{Sloan+15}
{Sloan}, G.~C., {Herter}, T.~L., {Charmandaris}, V., {et~al.} 2015,
  \href{http://dx.doi.org/10.1088/0004-6256/149/1/11}{\color{magenta}\aj},
  \href{https://ui.adsabs.harvard.edu/abs/2015AJ....149...11S}{\color{blue}149},
  \href{https://ui.adsabs.harvard.edu/abs/2015AJ....149...11S}{\color{blue}11}

\bibitem[{{Su} {et~al.}(2006){Su}, {Rieke}, {Stansberry}, {Bryden},
  {Stapelfeldt}, {Trilling}, {Muzerolle}, {Beichman}, {Moro-Martin}, {Hines},
  \& {Werner}}]{Su+06}
{Su}, K.~Y.~L., {Rieke}, G.~H., {Stansberry}, J.~A., {et~al.} 2006,
  \href{http://dx.doi.org/10.1086/508649}{\color{magenta}\apj},
  \href{https://ui.adsabs.harvard.edu/abs/2006ApJ...653..675S}{\color{blue}653},
  \href{https://ui.adsabs.harvard.edu/abs/2006ApJ...653..675S}{\color{blue}675}

\bibitem[{{Telesco} {et~al.}(2005){Telesco}, {Fisher}, {Wyatt}, {Dermott},
  {Kehoe}, {Novotny}, {Mari{\~n}as}, {Radomski}, {Packham}, {De Buizer}, \&
  {Hayward}}]{Telesco+05}
{Telesco}, C.~M., {Fisher}, R.~S., {Wyatt}, M.~C., {et~al.} 2005,
  \href{http://dx.doi.org/10.1038/nature03255}{\color{magenta}\nat},
  \href{https://ui.adsabs.harvard.edu/abs/2005Natur.433..133T}{\color{blue}433},
  \href{https://ui.adsabs.harvard.edu/abs/2005Natur.433..133T}{\color{blue}133}

\bibitem[{{Vacca} {et~al.}(2003){Vacca}, {Cushing}, \& {Rayner}}]{Vacca+03}
{Vacca}, W.~D., {Cushing}, M.~C., \& {Rayner}, J.~T. 2003,
  \href{http://dx.doi.org/10.1086/346193}{\color{magenta}\pasp},
  \href{https://ui.adsabs.harvard.edu/abs/2003PASP..115..389V}{\color{blue}115},
  \href{https://ui.adsabs.harvard.edu/abs/2003PASP..115..389V}{\color{blue}389}

\bibitem[{{Wahhaj} {et~al.}(2003){Wahhaj}, {Koerner}, {Ressler}, {Werner},
  {Backman}, \& {Sargent}}]{Wahhaj+03}
{Wahhaj}, Z., {Koerner}, D.~W., {Ressler}, M.~E., {et~al.} 2003,
  \href{http://dx.doi.org/10.1086/346123}{\color{magenta}\apjl},
  \href{https://ui.adsabs.harvard.edu/abs/2003ApJ...584L..27W}{\color{blue}584},
  \href{https://ui.adsabs.harvard.edu/abs/2003ApJ...584L..27W}{\color{blue}L27}

\bibitem[{{Weinberger} {et~al.}(2003){Weinberger}, {Becklin}, \&
  {Zuckerman}}]{Weinberger+03}
{Weinberger}, A.~J., {Becklin}, E.~E., \& {Zuckerman}, B. 2003,
  \href{http://dx.doi.org/10.1086/368065}{\color{magenta}\apjl},
  \href{https://ui.adsabs.harvard.edu/abs/2003ApJ...584L..33W}{\color{blue}584},
  \href{https://ui.adsabs.harvard.edu/abs/2003ApJ...584L..33W}{\color{blue}L33}

\bibitem[{{Wenger} {et~al.}(2000){Wenger}, {Ochsenbein}, {Egret}, {Dubois},
  {Bonnarel}, {Borde}, {Genova}, {Jasniewicz}, {Lalo{\"e}}, {Lesteven}, \&
  {Monier}}]{Wenger+2000}
{Wenger}, M., {Ochsenbein}, F., {Egret}, D., {et~al.} 2000,
  \href{http://dx.doi.org/10.1051/aas:2000332}{\color{magenta}\aaps},
  \href{https://ui.adsabs.harvard.edu/abs/2000A&AS..143....9W}{\color{blue}143},
  \href{https://ui.adsabs.harvard.edu/abs/2000A&AS..143....9W}{\color{blue}9}

\bibitem[{{Wooden} {et~al.}(2017){Wooden}, {Ishii}, \& {Zolensky}}]{Wooden+17}
{Wooden}, D.~H., {Ishii}, H.~A., \& {Zolensky}, M.~E. 2017,
  \href{http://dx.doi.org/10.1098/rsta.2016.0260}{\color{magenta}RSPTA},
  \href{https://ui.adsabs.harvard.edu/abs/2017RSPTA.37560260W}{\color{blue}375},
  \href{https://ui.adsabs.harvard.edu/abs/2017RSPTA.37560260W}{\color{blue}20160260}

\bibitem[{{Wozniakiewicz} {et~al.}(2012){Wozniakiewicz}, {Bradley}, {Ishii},
  {Brownlee}, {Kearsley}, {Burchell}, \& {Price}}]{Wozniakiewicz+12}
{Wozniakiewicz}, P.~J., {Bradley}, J.~P., {Ishii}, H.~A., {et~al.} 2012,
  \href{http://dx.doi.org/10.1088/2041-8205/760/2/L23}{\color{magenta}\apjl},
  \href{https://ui.adsabs.harvard.edu/abs/2012ApJ...760L..23W}{\color{blue}760},
  \href{https://ui.adsabs.harvard.edu/abs/2012ApJ...760L..23W}{\color{blue}L23}

\bibitem[{{Wright} {et~al.}(2010){Wright}, {Eisenhardt}, {Mainzer}, {Ressler},
  {Cutri}, {Jarrett}, {Kirkpatrick}, {Padgett}, {McMillan}, {Skrutskie},
  {Stanford}, {Cohen}, {Walker}, {Mather}, {Leisawitz}, {Gautier}, {McLean},
  {Benford}, {Lonsdale}, {Blain}, {Mendez}, {Irace}, {Duval}, {Liu}, {Royer},
  {Heinrichsen}, {Howard}, {Shannon}, {Kendall}, {Walsh}, {Larsen}, {Cardon},
  {Schick}, {Schwalm}, {Abid}, {Fabinsky}, {Naes}, \& {Tsai}}]{Wright+10}
{Wright}, E.~L., {Eisenhardt}, P. R.~M., {Mainzer}, A.~K., {et~al.} 2010,
  \href{http://dx.doi.org/10.1088/0004-6256/140/6/1868}{\color{magenta}\aj},
  \href{https://ui.adsabs.harvard.edu/abs/2010AJ....140.1868W}{\color{blue}140},
  \href{https://ui.adsabs.harvard.edu/abs/2010AJ....140.1868W}{\color{blue}1868}

\bibitem[{{Xu} {et~al.}(2014){Xu}, {Jura}, {Koester}, {Klein}, \&
  {Zuckerman}}]{Xu+14}
{Xu}, S., {Jura}, M., {Koester}, D., {et~al.} 2014,
  \href{http://dx.doi.org/10.1088/0004-637X/783/2/79}{\color{magenta}\apj},
  \href{https://ui.adsabs.harvard.edu/abs/2014ApJ...783...79X}{\color{blue}783},
  \href{https://ui.adsabs.harvard.edu/abs/2014ApJ...783...79X}{\color{blue}79}

\bibitem[{{Zeidler} {et~al.}(2015){Zeidler}, {Mutschke}, \&
  {Posch}}]{Zeidler+15}
{Zeidler}, S., {Mutschke}, H., \& {Posch}, T. 2015,
  \href{http://dx.doi.org/10.1088/0004-637X/798/2/125}{\color{magenta}\apj},
  \href{https://ui.adsabs.harvard.edu/abs/2015ApJ...798..125Z}{\color{blue}798},
  \href{https://ui.adsabs.harvard.edu/abs/2015ApJ...798..125Z}{\color{blue}125}

\end{thebibliography}
\end{document}